\documentclass[aps, pra, superscriptaddress, floatfix, reprint]{revtex4-1}
\usepackage{graphicx}
\usepackage{dcolumn}
\usepackage{bm}
\usepackage{amsfonts}
\usepackage{amsmath}
\usepackage{amssymb}
\usepackage{color}
\usepackage[normalem]{ulem}
\usepackage[percent]{overpic}

\usepackage[colorlinks=true,citecolor=blue]{hyperref}
\hypersetup{colorlinks=true,citecolor=blue,linkcolor=red,urlcolor=blue}

\def\be{\begin{equation}}
\def\ee{\end{equation}}

\begin{document}

\title{Rotons and Bose condensation in Rydberg-dressed Bose Gases}

\author{Iran Seydi}
	\affiliation{Department of Physics, Institute for Advanced Studies in Basic Sciences (IASBS), Zanjan 45137-66731, Iran}
\author{Saeed H. Abedinpour}
\email{abedinpour@iasbs.ac.ir}
	\affiliation{Department of Physics, Institute for Advanced Studies in Basic Sciences (IASBS), Zanjan 45137-66731, Iran}
	\affiliation{Research Center for Basic Sciences \& Modern Technologies (RBST), Institute for Advanced Studies in Basic Sciences (IASBS), Zanjan 45137-66731, Iran}
		\affiliation{School of Nano Science, Institute for Research in Fundamental Sciences (IPM), Tehran 19395-5531, Iran}
\author{Robert E. Zillich}
	\affiliation{Institute for Theoretical Physics, Johannes Kepler University, Altenbergerstrasse 69, 4040 Linz, Austria}
\author{Reza Asgari}
	\affiliation{School of Physics, Institute for Research in Fundamental Sciences (IPM), Tehran 19395-5531, Iran}
	\affiliation{School of Nano Science, Institute for Research in Fundamental Sciences (IPM), Tehran 19395-5531, Iran}
\author{B. Tanatar}
	\affiliation{Department of Physics, Bilkent University, Bilkent, 06800 Ankara, Turkey}

\date{\today}

\begin{abstract}
We investigate the ground-state properties and excitations of Rydberg-dressed bosons in both three and two dimensions, using the hypernetted-chain Euler-Lagrange approximation, which accounts for correlations and thus goes beyond the mean field approximation. The short-range behavior of the pair distribution function signals the instability of the homogeneous system towards the formation of droplet crystals at strong couplings and large soft-core radius.
This tendency to spatial density modulation coexists with off-diagonal long-range order.
The contribution of the correlation energy to the ground-state energy is significant at large coupling strengths and intermediate values of the soft-core radius while for a larger soft-core radius the ground-state energy is dominated by the mean-field (Hartree) energy.
We have also performed path integral Monte Carlo simulations to verify the performance of our hypernetted-chain Euler-Lagrange results in three dimensions.
In the homogeneous phase, the two approaches are in very good agreement. Moreover, Monte Carlo simulations predict a first-order quantum phase transition from a homogeneous superfluid phase to the quantum droplet phase with face-centered cubic symmetry for Rydberg-dressed bosons in three dimensions.
\end{abstract}
\maketitle

\section{Introduction}
Rydberg systems consisting of atoms with a highly excited electron~\cite{loewJPhysB12} have attracted a lot of interest in recent years through studying a variety of quantum many-body~\cite{henkel2012supersolid,PhysRevLett.105.135301,Cinti}, quantum information~\cite{RevModPhys.82.2313,ripkaScience18}, quantum simulation~\cite{weimer2010rydberg,browaeys2016experimental},
and polaron~\cite{camargoPRL18} problems.
Rydberg atoms in the blockade regime, in particular, are expected to become important tools for quantum information as the manipulation of the entanglement of two or more atoms in these systems are very feasible~\cite{WALKER201281,engelPRL18}.
In this regime, a Rydberg atom shifts the energy levels of its neighboring atoms. This effect results from the strong interaction between a Rydberg atom and its surrounding ground-state atoms, and therefore a single Rydberg atom can block the excitation of other atoms in its neighborhood~\cite{PhysRevLett.114.203002}.

Rydberg atoms possess very strong van der Waals interactions, but short lifetimes of excited atoms would be an obstacle in experiments.
A solution around this problem is to weakly dress the ground state with a small fraction of the Rydberg state, which results in several orders of magnitude enhancement of the lifetime~\cite{PhysRevLett.104.195302, browaeys2016experimental, zeiher2016many}.
The effective Rydberg-dressed interaction potential is almost constant at short inter-particle distances and has a van der Waals i.e., $1/r^{6}$-tail at large separations~\cite{browaeys2016experimental}.
Several novel quantum phases have been predicted for Rydberg-dressed quantum gases, such as the super-solid phase~\cite{PhysRevLett.105.135301, Cinti, henkel2012supersolid, macri2014ground,  PhysRevA.84.063621}, topological superfluidity~\cite{PhysRevA.90.013631}, metallic quantum solid phase~\cite{PhysRevLett.117.035301}, density waves~\cite{PhysRevA.96.053611}, and roton excitations~\cite{PhysRevLett.104.195302}.
A Rotating quasi-two-dimensional Rydberg-dressed Bose-Einstein condensate (BEC) has been studied by Henkel and coauthors~\cite{henkel2012supersolid} by means of quantum Monte Carlo simulations and mean field calculations. They have predicted s superfluid phase for slow rotations, as well as a competition between the super-solid crystal and a vortex lattice for rapid rotations.
The zero-temperature phase diagram of two-dimensional bosons with a finite-range soft-core interaction has also been studied in the framework of the path-integral Monte Carlo method by Cinti \emph{et al.}~\cite{Cinti}. 
Depending on the particle density and interaction strength, they found superfluid, supersolid and different solid phases. For small particle densities, they have predicted a defect induced supersolid phase~\cite{Cinti}.
On the experimental side, supersolidity in an optical lattice composed of strongly correlated Rydberg dressed bosons has been explored~\cite{li2018supersolidity}.

In this work, we investigate the effects of many-body correlations on the ground-state properties of a single-component gas of Rydberg dressed bosons (RDB) in both three and two dimensions (abbreviated as 3D and 2D, respectively), within the framework of the hypernetted-chain Euler-Lagrange (HNC-EL) approximation.
We have obtained several ground-state quantities, as well as the excitation spectra which for strong coupling and
large soft core radius feature pronounced rotons. Roton softening has been suggested in mean field calculations
as a mechanism of destabilizing the homogeneous phase and leading to a crystalline phase~\cite{PhysRevLett.104.195302}.
We have performed path integral Monte Carlo (PIMC) simulations for a 3D gas of RDB at selected system parameters and found very good agreement between PIMC and HNC-EL results in the homogeneous superfluid phase.
The PIMC results suggest a first-order transition from a homogenous superfluid phase to a face-centered cubic (FCC) lattice formed of quantum droplets, in agreement with the mean field calculations in Ref.~\onlinecite{PhysRevLett.104.195302}.

The rest of this paper is organized as follows.
We begin with a description of our theoretical formalism in Sec.~\ref{sec:theory}, followed by the details of the HNC-EL approximation for obtaining the static structure factor and pair distribution function (PDF) in subsection~\ref{sec:HNC} and the method for calculating the one-body density matrix as well as the momentum distribution function within the HNC-EL formalism in subsection~\ref{sec:1BDM}.
In Sec.~\ref{sec:results}, we report our numerical results for different quantities obtained within the HNC-EL approximation.
The details of PIMC simulations and the comparison between its results with HNC-EL results are presented in Sec.~\ref{sec:PIMC}.
Finally, in Sec.~\ref{sec:summ} we summarize our main findings.

\section{Model and theoretical formalism}\label{sec:theory}
We consider a homogeneous single component gas of RDB of mass $m$, in both three and two dimensions, where each atom is weakly coupled to its s-wave Rydberg state by an off-resonant two-photon transition via an intermediate state. The Hamiltonian is thus given by
\begin{equation}
  H = -{\hbar^2\over 2m}\sum_i\nabla^2_i + \sum_{i<j}v_{\rm RD}(|{\bf r}_i-{\bf r}_j|).
\label{eq:Ham}
\end{equation}
Dressed-state atoms interact with each other through the following repulsive soft-core potential~\cite{PhysRevLett.104.195302}
\begin{equation}\label{eq:Ryd_int}
v_{\rm RD}(r)= \frac{U}{1 + (r/R_{c})^6}.
\end{equation}
Here, $U\equiv ({\Omega}/{2\Delta})^{4} |C_{6}| / R_{c}^{6}$ and $R_{c}  \equiv (C_{6} / 2 \hbar \Delta)^{1/6}$ are the interaction strength and the averaged soft-core radius, respectively, where $\Omega$, $\Delta<0$, and $C_{6}<0$ are the effective Raman coupling, the red detuning, and the averaged van der Waals coefficient, respectively.

By introducing $k_0^{3D} = (6{\pi}^{2}n)^{\frac{1}{3}}$ and $k_0^{2D} = \sqrt{4 \pi n}$, respectively, in three dimensional and two dimensional systems, where $n$ is the corresponding average particle density of bosons, the RDB gas at zero temperature would be characterized by only two dimensionless parameters, namely
the dimensionless soft-core radius $\tilde{R_c} = {R_c}{k_0}$, and the dimensionless coupling constant $\tilde{U} = U / {\varepsilon _0}$, where ${\varepsilon _0} = \hbar^2 k_{0}^2 / (2m)$.

The bare potential~\eqref{eq:Ryd_int} has an almost constant value $U$ at small distances $r << R_c$ and approaches zero as $1/r^{6}$ for $r >> R_c$.
While the Rydberg-dressed interaction is purely repulsive in real space, its Fourier transform has a negative minimum at a finite wave vector 
$q_{\rm min} \approx 5 / R_{c}$~\cite{macri2014ground, PhysRevLett.117.035301, PhysRevA.96.053611}.

\subsection{Hypernetted-chain approximation}\label{sec:HNC}
By choosing the chemical potential as the zero of energy, a formally exact zero-energy scattering equation for the pair distribution function $g(r)$ of a homogeneous Bose system can be written within the hypernetted-chain Euler-Lagrange approximation~\cite{feenberg1969theory,Fabrocini2002book}
\begin{equation}\label{eq:diff_g}
\left[ -\frac{\hbar^2}{m} \nabla^{2} + W_{\rm eff}(r)\right]\sqrt{g(r)} = 0.
\end{equation}
Here, $W_{\rm eff}(r) = v_{\rm RD}(r) + W_{\rm B}(r)$ is the effective scattering potential consisting of the bare interaction $v_{\rm RD}(r)$ and an induced interaction $W_{\rm B}(r)$ accounting for many-body effects.  
For a homogeneous system, the pair distribution function is given by 
\begin{equation}
g(r - r^{\prime}) = \frac{N-1}{n^2} \int \mathrm{d}\textbf{r}_{3} ...\mathrm{d}\textbf{r}_{N}|\Psi (\textbf{r},\textbf{r}' ,\textbf{r}_{3} , ... , \textbf{r}_{N})|^{2},
\end{equation}
where $\Psi (\textbf{r}_{1}, \textbf{r}_{2} , ... , \textbf{r}_{N})$ is the many-body wave function of the system normalized to the total number of particles $N= \int \mathrm{d}\textbf{r}_{1} ...\mathrm{d}\textbf{r}_{N} |\Psi (\textbf{r}_{1}, \textbf{r}_{2} , ... , \textbf{r}_{N})|^{2}$.
The PDF $g(r)$ is defined such that $ng(r)\Omega_D r^{D-1} \mathrm{d}r$, with $\Omega_2 = 2\pi$
and $\Omega_3= 4\pi$, is the average number of particles inside a shell of radius $r$ and thickness $\mathrm{d}r$ centered on the particle at the origin and therefore it is a positive-definite function.
The normalization of $g(r)$ is chosen so that $ g(r\rightarrow \infty) \rightarrow 1$, meaning that correlations between two particles vanishes at large separations~\cite{mahan1990many,giuliani2005quantum}, and in a noninteracting homogeneous Bose gas $g_0(r)=1$.

Indeed, in the limit of vanishing density, $W_{\rm B}(r)$ vanishes and Eq.~(\ref{eq:diff_g}) becomes the Schr\"odinger equation for two-body scattering at zero energy.  $W_{\rm B} (r)$, at the level of the so-called HNC-EL/0 approximation~\cite{Fabrocini2002book,{KROTSCHECK19931}}, is given in momentum space by
\be\label{eq:W_B}	
W_{\rm B}(q) = -\frac{\varepsilon_q}{2n}\left[2S(q) + 1\right] \left[\frac{S(q) - 1}{S(q)}\right]^{2},
\ee
where ${\varepsilon_q}=\hbar^2q^2/(2m)$ is the free particle dispersion and the static structure factor $S(q)$ is related to the $g(r)$ as $S(q) = 1 + n {\rm FT}[g(r) - 1]$, with FT[$f(r)$]=$\int \mathrm{d}{\bf r} f({\bf r}) e^{i{\bf k}\cdot {\bf r}}$ being a short hand notation for the Fourier transform.
In principle, Eqs.~\eqref{eq:diff_g} and \eqref{eq:W_B}, could be solved self-consistently but technically it is more convenient to invert the zero-energy scattering equation~\eqref{eq:diff_g} to obtain the effective particle-hole interaction
\begin{equation}\label{eq: V_{ph}}
V_{\rm ph}(r) = g(r) W_{\rm eff}(r) -W_{\rm B}(r) + \frac{\hbar^2}{m} \left| \nabla \sqrt{g(r)}\right|^2,
\end{equation}
whose Fourier space expression is defined in terms of $S(q)$ as
\begin{equation}\label{eq:static S}
S(q) = \frac{1}{\sqrt{1 + 2 n V_{\rm ph}(q) / \varepsilon_q}}.
\end{equation}
Now, Eqs.~\eqref{eq:W_B}-\eqref{eq:static S} form a closed set of equations, which can be solved in a self-consistent manner with a reasonable first guess for the static structure factor. The self-consistent process is repeated until convergence is reached \cite{PhysRevA.86.043601}.
We have used the HNC-EL/0 approximation, which corresponds to a Jastrow-Feenberg ansatz for the many-body wave function containing only two-body correlations but no three-body and higher order correlations and in addition neglecting the so-called elementary diagrams.  We expect contributions beyond the HNC-EL/0 approximation to be small at weak couplings and mainly quantitative corrections at intermediate and strong couplings. This will become clear from the comparison between our HNC-EL/0 and PIMC results in section~\ref{sec:PIMC}.

%

\subsection{One-body density matrix and momentum distribution} \label{sec:1BDM}
Once the static structure factor is obtained from the solution of self-consistent HNC-EL/0 equations, we can calculate several important quantities such as the one-body density matrix (OBDM), the condensate fraction, and the momentum distribution function within the HNC-EL/0 formalism. For a homogenous system, the OBDM is given by
\begin{equation}
\rho(r) = \int \mathrm{d}\textbf{r}_{2} ...\mathrm{d}\textbf{r}_{N} \Psi ^{*}(\textbf{r}, \textbf{r}_{2} , ... , \textbf{r}_{N}) \Psi (\textbf{0}, \textbf{r}_{2} , ... , \textbf{r}_{N}),
\end{equation}
which at the origin gives the average density $\rho(0) = n$, while for long distances it is a measure of the off-diagonal long-range order (ODLRO), i.e., $\rho(r \rightarrow \infty) =n  n_{0}$, where $n_{0}$ is the Bose-Einstein condensation (BEC) fraction.
Within the HNC-EL/0 formalism, i.e. neglecting elementary diagrams,
the OBDM is given by~\cite{PhysRevB.31.7022,PhysRevA.86.043601}
\begin{equation}
\rho(r) = n n_{0} e^{N_{ww}(r)},
\end{equation}
where the Fourier transform of the nodal function $N_{ww}(r)$ is given by
\begin{equation}
N_{ww}(q) = \left[S_{wd}(q) - 1 \right] \left[S_{wd}(q) - 1 -  N_{wd}(q) \right].
\end{equation}
Here, $ S_{wd}(q)$ and $ N_{wd}(q)$ are, respectively, obtained from the solutions of the following coupled equations
\begin{equation}\label{eq:N_w}
N_{wd}(q) = \left[ S_{wd}(q) - 1 \right] \left[ S(q) - 1 -  N(q) \right],
\end{equation}
and
\begin{equation}\label{eq:S_w}
S_{wd}(q) = 1 + n {\rm FT}\left[ g_{wd}(r) - 1 \right],
\end{equation}
with
\begin{equation}\label{eq:g_w}
g_{wd}(r) = f(r) e^{N_{wd}(r)}.
\end{equation}
Here, $N(q) = \left[S(q) -1\right]^2/S(q)$
is the nodal function, and $f(r) = \sqrt{g(r) exp \left[ -N(r) \right] }$ is the correlation function.
Now, we can solve Eqs.~\eqref{eq:N_w}-\eqref{eq:g_w} self-consistently and then obtain the condensation fraction from
\begin{equation}
n_{0} = \exp (2 R_{w} - R_{d}),
\end{equation}
where
\begin{equation}
\begin{split}
R_{w} = &n \int \mathrm{d}{\textbf{r}} \left[ g_{wd}(r) - 1 - N_{wd} (r) \right]  \\
&- \frac{n}{2} \int \mathrm{d}{\textbf{r}} \left[ g_{wd}(r) - 1 \right] N_{wd} (r),
\end{split}
\end{equation}
and
\begin{equation}
\begin{split}
R_{d} = &n \int \mathrm{d}{\textbf{r}} \left[ g(r) - 1 - N(r) \right]  \\
 &- \frac{n}{2} \int \mathrm{d} \textbf{r} \left[g(r)  - 1 \right] N (r).
\end{split}
\end{equation}
Finally, the momentum distribution function could be obtained from the Fourier transformation of the OBDM
\begin{equation}
n(q) = n n_{0} (2\pi)^{2} \delta (\textbf{q}) + n {\rm FT} \left[\rho(r) / n - n_{0}\right] .
\end{equation}

\section{Results and Discussions}\label{sec:results}
In this section, we present our numerical results obtained from the HNC-EL/0 formalism for different ground state properties of homogenous Rydberg-dressed Bose gases in two and three dimensions.

\subsection{Static structure factor and excitation spectrum}
\begin{figure}
\centering
\includegraphics[width=\linewidth]{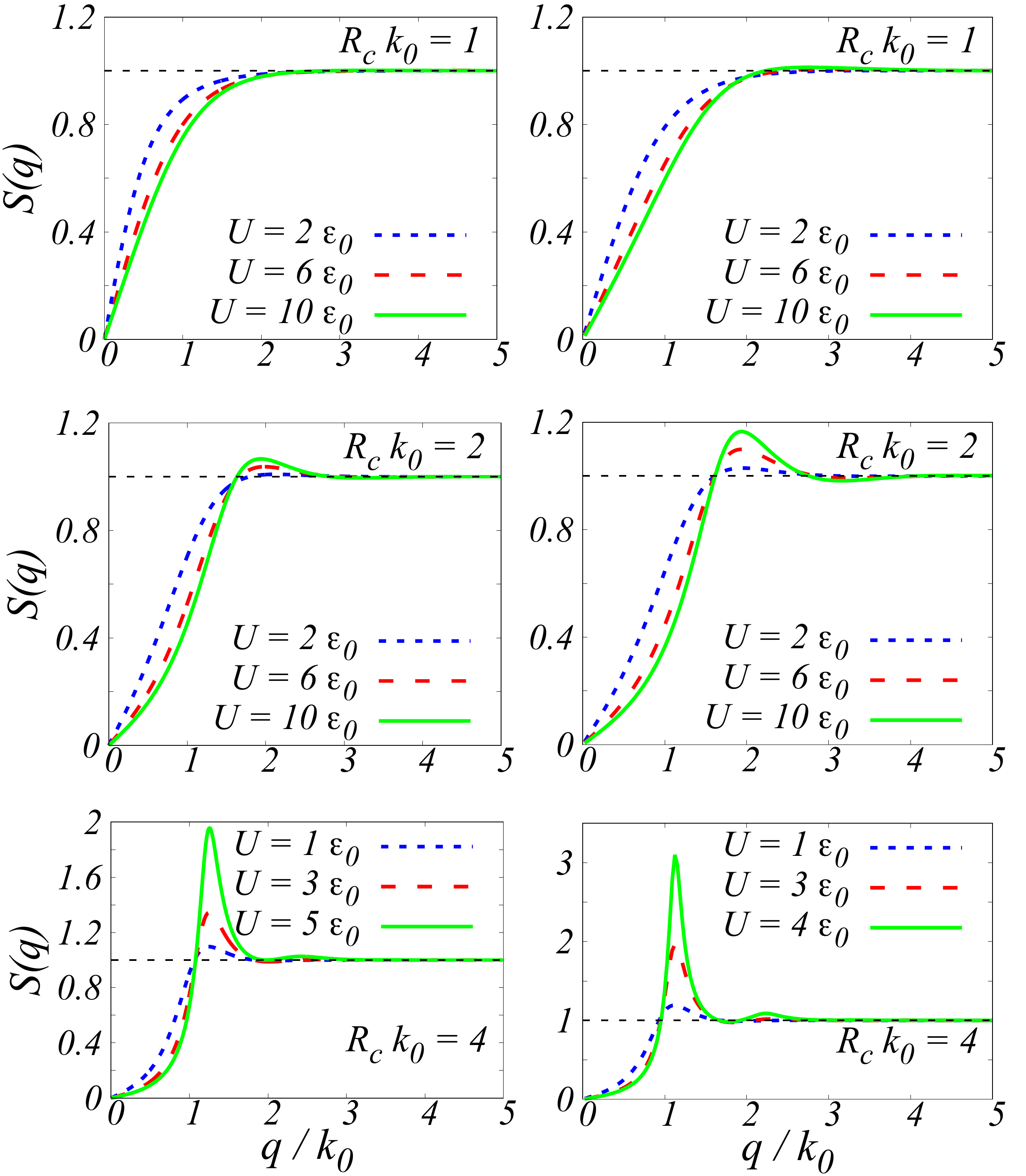}
\caption{The static structure factor $S(q)$ versus $q / k_0$ obtained within the HNC-EL/0 approximation for several values of $\tilde{R}_{c}$ and $\tilde{U}$, for 3D (left) and 2D (right) Rydberg-dressed Bose gas.
\label{fig:Sq}}
\end{figure}

Fig.~\ref{fig:Sq} shows our results for the static structure factor of 3D and 2D Rydberg-dressed Bose gases at different values of $\tilde{R_{c}}$ and $\tilde U$. When the strength of the coupling constant is increased, correlations get stronger and the height of the main peak in $S(q)$ increases.
For similar values of ${\tilde U}$ and ${\tilde R}_c$, the main peak of the structure factor in a 2D system is more pronounced than in a 3D system. This is expected, as the correlations are generally stronger in lower dimensions.

\begin{figure}
\centering
\includegraphics[width=1.0\linewidth]{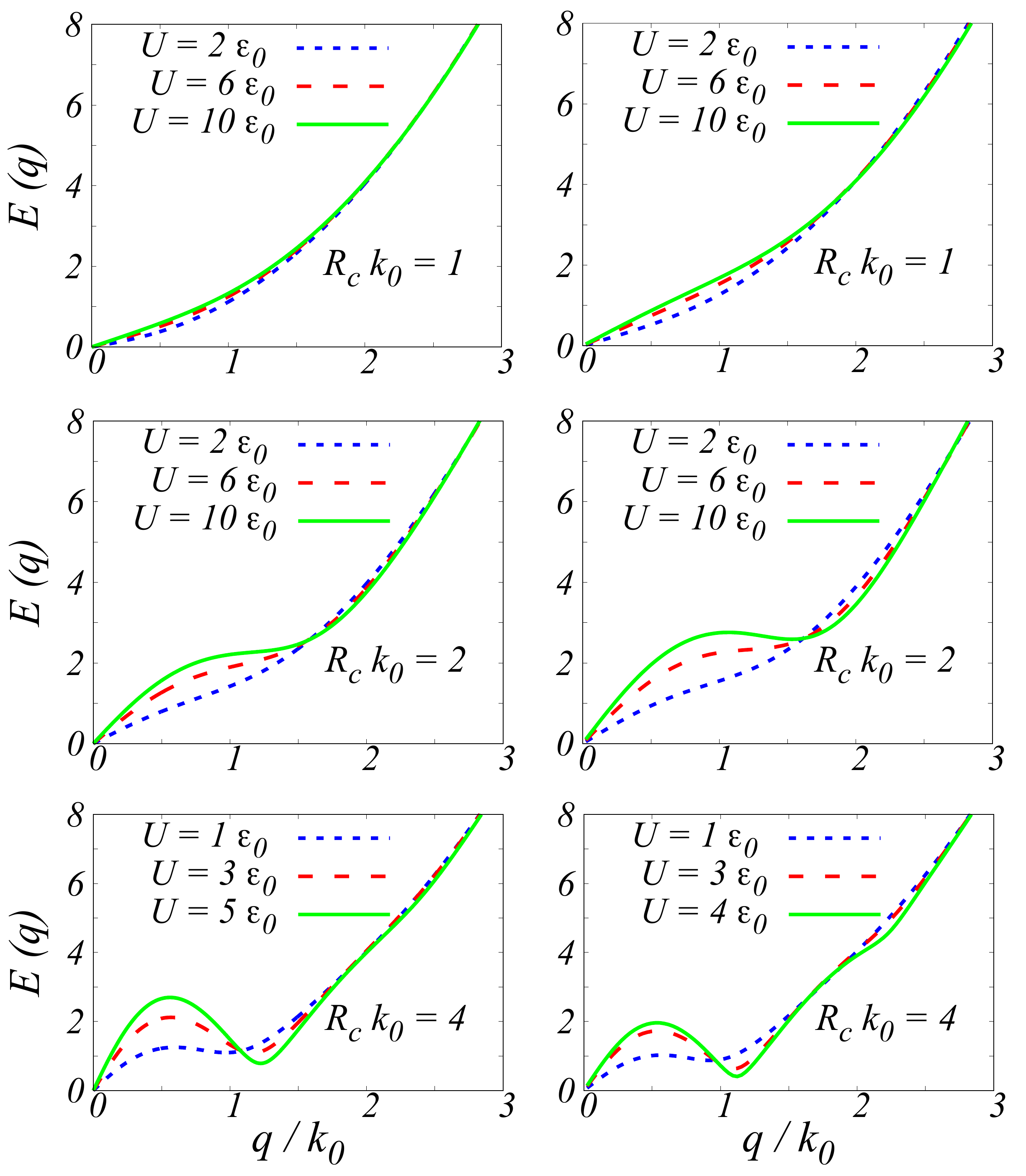}
\caption{The excitation spectrum $E(q)$ of a 3D (left) and 2D (right) RDB gas (in units of $\varepsilon_0$) versus $q / k_0$ for different values of $\tilde{R}_{c}$ and $\tilde{U}$, obtained within the HNC-EL/0 approximation.
\label{fig:E_k}}
\end{figure}

An upper bound for the excitation spectrum can be obtained from the Bijl-Feynman (BF) expression $E(q)= \varepsilon_q/S(q)$~\cite{{PhysRev.102.1189},{PhysRev.94.262}}.
In Fig.~\ref{fig:E_k} we show the excitation spectrum $E(q)$ of 3D and 2D Rydberg-dressed Bose gases.
In all cases, the spectrum has a linear behavior at small $q$, as expected for a uniform gas of interacting bosons. 
In single component Bose gases the BF approximation captures this small $q$ behavior very well.
For large $q$, the dispersion becomes parabolic because the static structure factor tends to unity at large wave vectors, and the BF spectrum becomes that of free particles~\cite{annett2004superconductivity}.
In the intermediate and strong coupling regimes and for small and intermediate soft-core radii, the excitation spectra $E(q)$ has a roton-maxon form, that is a local maximum at $q_{\rm maxon}$ is followed by a local minimum at $q_{\rm roton}$.
We caution that beyond the linear regime of the dispersion, the BF approximation overestimates excitation energies and furthermore neglects spectral broadening, which becomes quite noticeable for strong interactions.  Improved methods beyond the BF approximation, as discussed in the conclusions, are beyond the scope of the present work.

Increasing the interaction strength at a fixed soft-core radius, as the main peak of structure factor becomes more pronounced, the numerical convergence of HNC-EL/0 equations becomes very difficult.
The vanishing of the BF roton energy, which originates from the divergence of the static structure factor,
would signal the instability of a homogeneous superfluid towards density modulated phases with wavelength
$\lambda=2\pi/q_{\rm roton}$.
Such an instability corresponds to a second-order phase transition, but as we will see in section \ref{sec:PIMC}, quantum Monte Carlo simulations predict a first-order fluid to solid phase transition which precedes such an
instability. Since we apply the HNC-EL/0 method of homogeneous systems, the HNC-EL/0 results beyond the
phase transition are only metastable. Generalizations of HNC-EL for lattice symmetries have recently been
presented for 1D in Ref.~\cite{panholzerJLTP17}.

\subsection{Pair distribution function and effective interaction}
\begin{figure}
\centering
\includegraphics[width=1.0\linewidth]{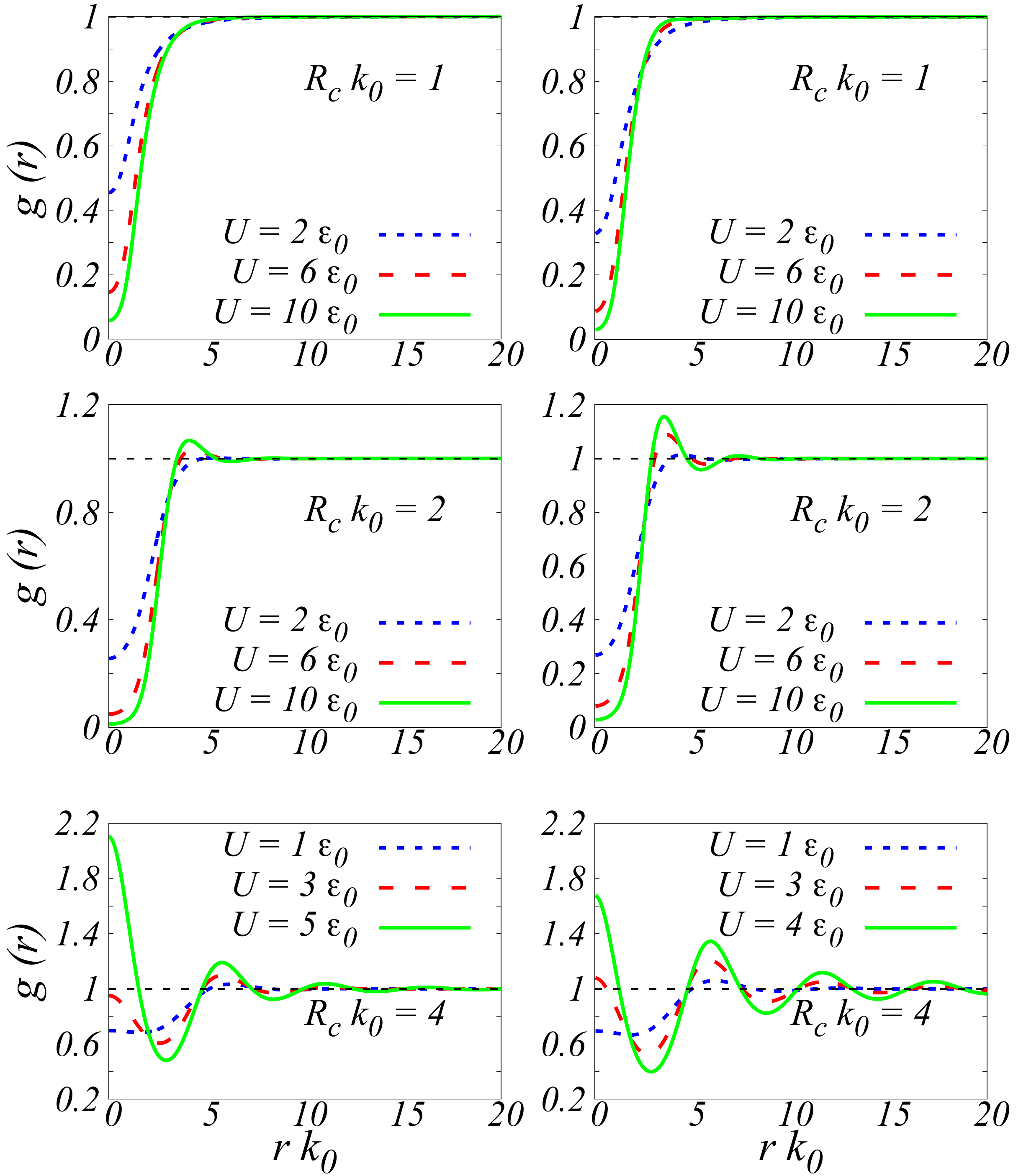}
\caption{The pair distribution function $g(r)$ versus $r k_0$, obtained within the HNC-EL/0 approximation at several values of $\tilde{R}_{c}$ and $\tilde{U}$ for  3D (left) and 2D (right) gas of RDB.
\label{fig:gr}}
\end{figure}

\begin{figure}
\centering
\begin{tabular}{c}
\includegraphics[width=0.8\linewidth]{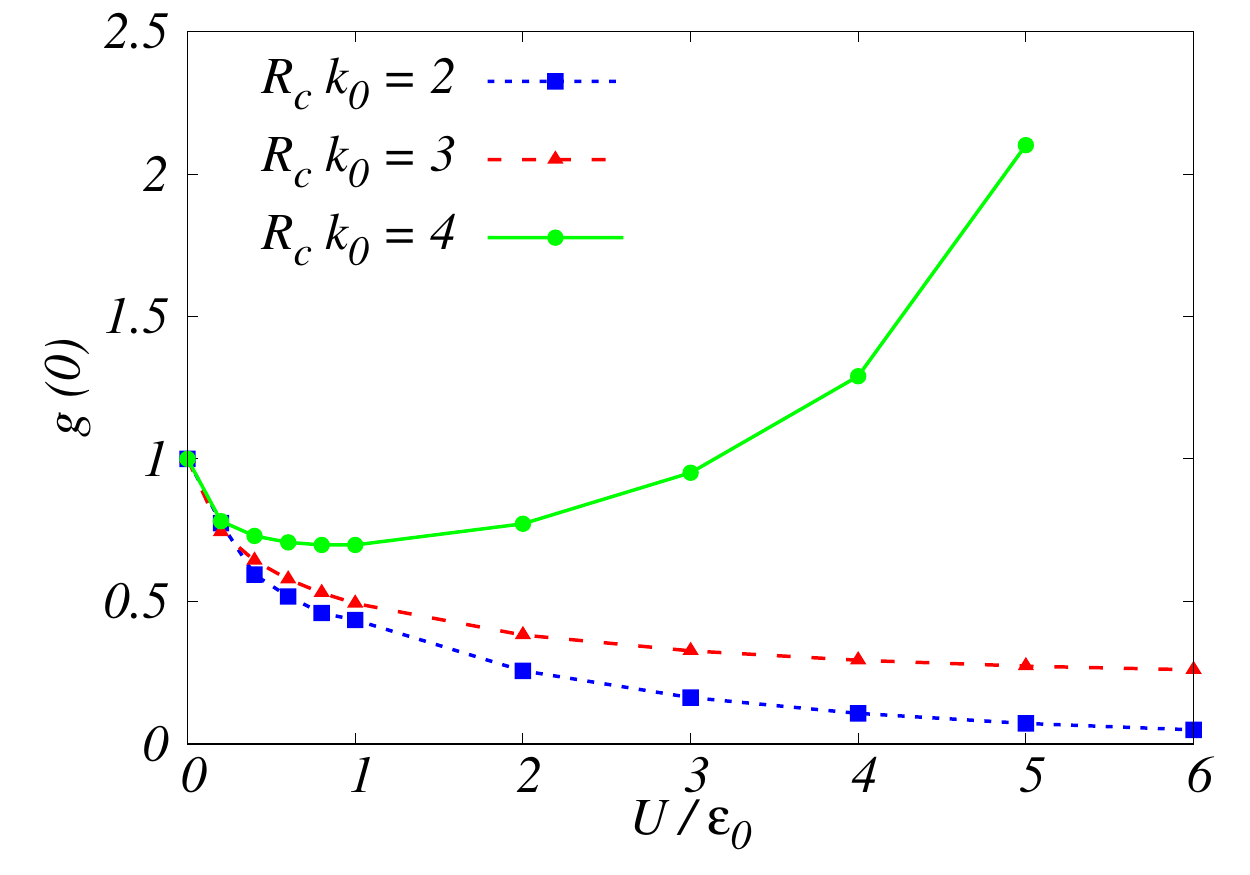}\\
\includegraphics[width=0.8\linewidth]{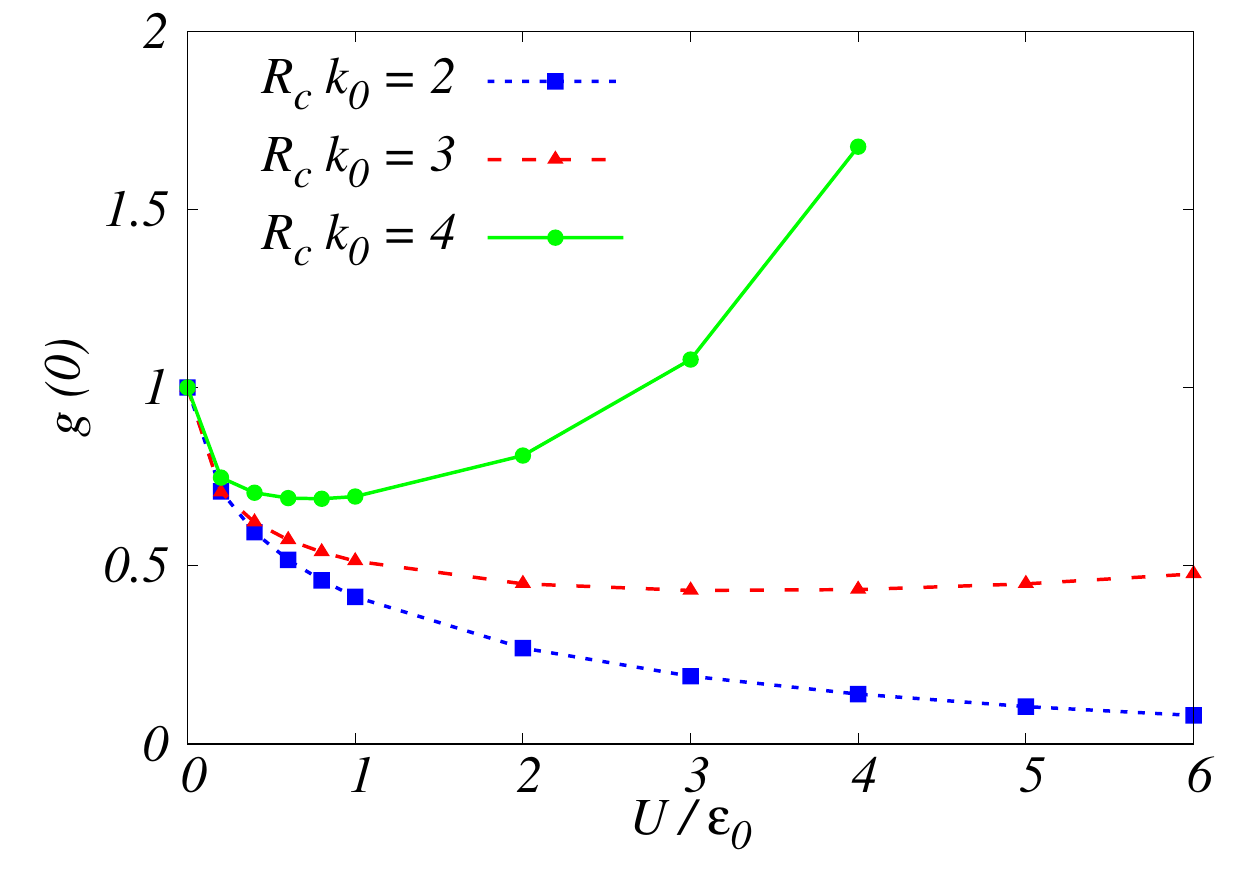}
\end{tabular}
\caption{The on-top-value of the pair distribution function $g(0)$ versus the interaction strength $U/\varepsilon_0$, for different values of the soft-core radius in 3D (top) and 2D (bottom) gas of RDB.
Note that, for ${\tilde R}_c=4$, the HNC-EL/0 equations fail to converge at large values of the interaction strength.
\label{fig:g0}}
\end{figure}

We present our results for the pair distribution function of 3D and 2D RDB gases at different values of $\tilde{R}_{c}$ and $\tilde{U}$ in Fig.~\ref{fig:gr}.
For small values of the soft-core radius $\tilde{R}_{c}$, the probability for particles to coincide spatially, i.e. $g(r=0)$, decreases with increasing interaction strength ${\tilde U}$. This indicates the formation of a correlation hole around each particle~\cite{giuliani2005quantum}, due to the repulsive interaction between particles.
However, an interesting feature emerges at larger values of $\tilde{R}_{c}$ (see, the bottom panels in Fig.~\ref{fig:gr}), where after an initial decrease, $g(0)$ starts increasing for larger interaction
strengths $\tilde{U}$ and eventually exceeds one. This means there is a positive correlation for particles
to assume the same position in space, i.e. they tend to cluster up.
The PDF exhibits slowly decaying oscillations in this regime.
The behavior of $g(0)$ as function of $\tilde{U}$ for different values of $\tilde{R}_{c}$ is
summarized in Fig.~\ref{fig:g0} for 3D (top panel) and 2D (bottom panel).  We note that the
probability for two particles to meet is given by $n^2 g(r=0)$, and may be used to estimate the three-particle decay rate, which in the Kirkwood superposition approximation would be $n^3g(0)^3$.

\begin{figure}
\centering
\includegraphics[width=1.0\linewidth]{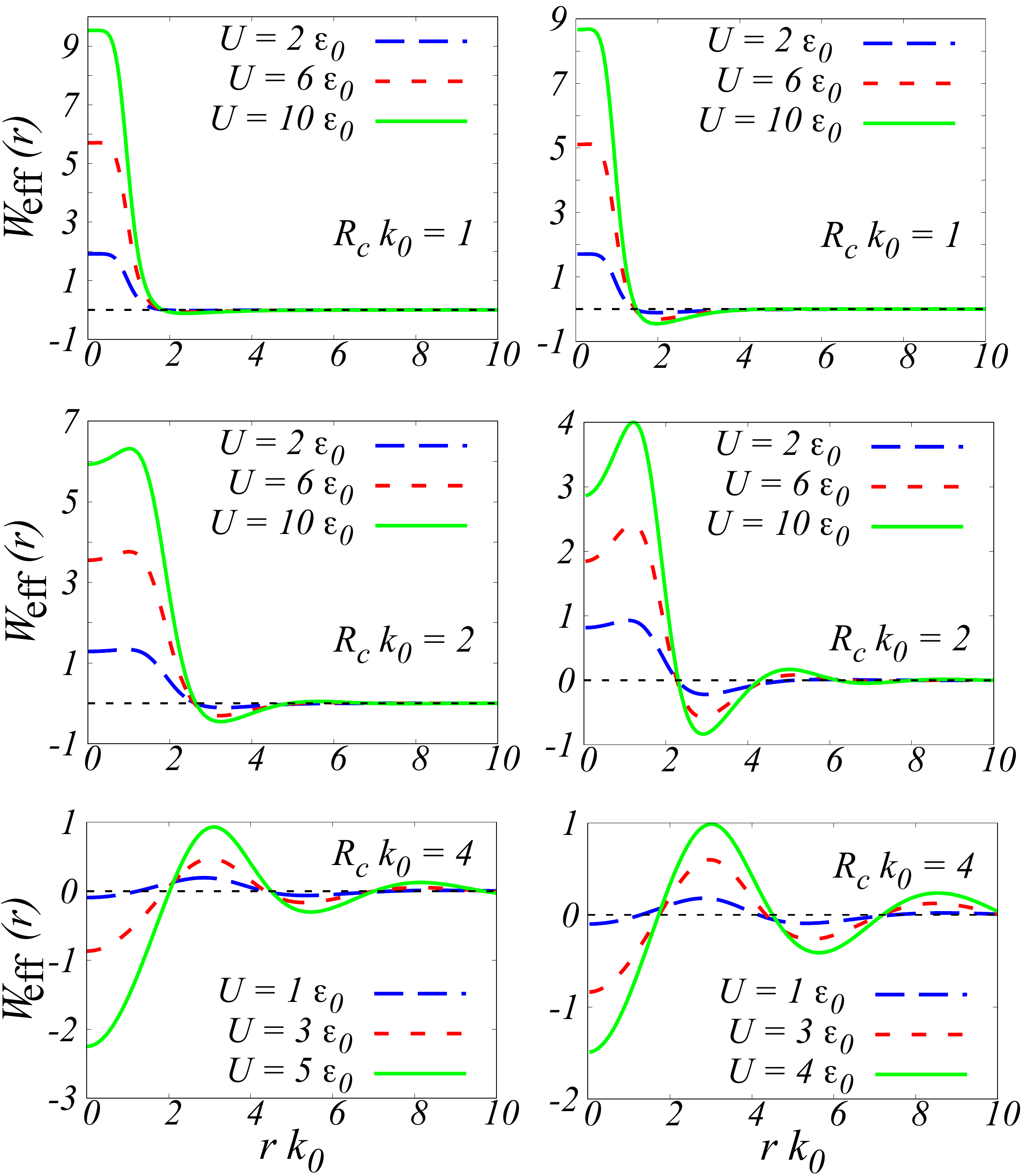}
\caption{The effective interaction $W_{\rm eff}(r)$ (in units of $\varepsilon_0$) versus $r k_0$ obtained  within the HNC-EL/0 approximation for different values of $\tilde{R}_{c}$ and $\tilde{U}$ in 3D (left) and 2D (right) gas of RDB.
\label{fig:Weff}}
\end{figure}

This peculiar behavior of PDF could be understood from the effective interaction $W_{\rm eff}(r)$, which is illustrated in Fig.~\ref{fig:Weff}. While the effective interaction at small $r$ is repulsive for small and intermediate values of the soft-core radius, for larger values of ${\tilde R}_c$ it becomes a strongly oscillating function of $r$ and attractive at small distances (see, the bottom panels in Fig.~\ref{fig:Weff}). This behavior can signal that the homogeneous Bose gas becomes soft against both droplet formation -- indicated by the increased $g(0)$ --
and forming density waves -- indicated by the long range of oscillations. Hence the behavior of $g(r)$ suggest the
RDB gas becomes unstable against forming a droplet crystal~\cite{{PhysRevLett.105.135301},{Cinti}}.
Again, due to stronger correlations at lower dimensions, the tendency to establish long range order in a 2D system shows up at smaller values of the coupling constant in comparison with a 3D system.

\begin{figure}
\includegraphics[width=\linewidth]{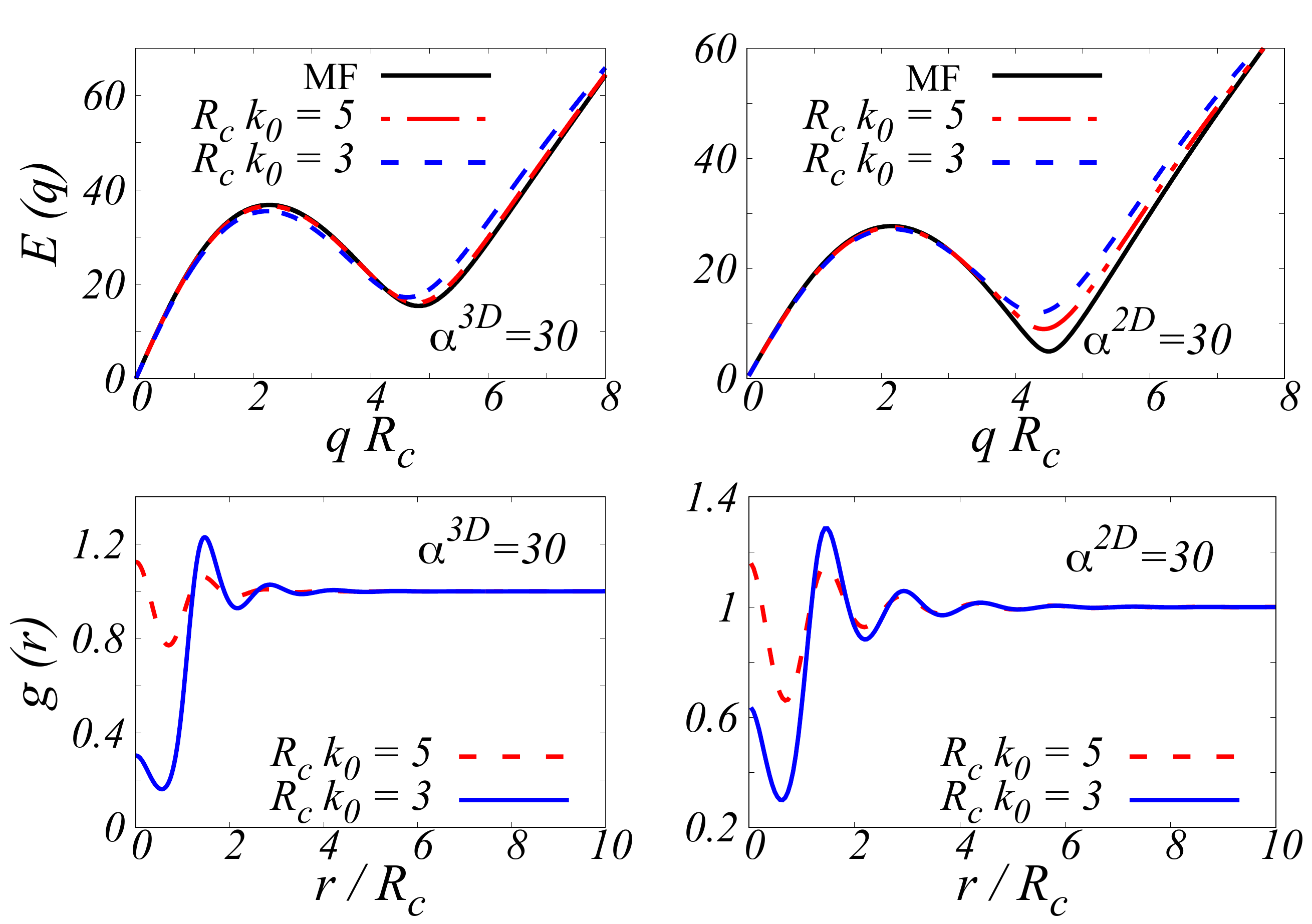}
\caption{Top panels: Comparison between the mean-filed (solid black) and HNC-EL/0 excitation spectrum $E(q)$ [in units of $\hbar^2/(m R_c^2)$] of a 3D (left) and 2D (right) RDB gas  versus $q R_c$, for different values of $\tilde{R}_{c}$ at $\alpha=30$. Bottom panels: HNC-EL/0 results for the pair correlation function $g(r)$ of a 3D (left) and 2D (right) RDB gas versus $r / R_c$ for different values of $\tilde{R}_{c}$ and for $\alpha=30$.
\label{fig:E_MF}}
\end{figure}

In the weak coupling regime 
the quasiparticle excitation spectrum can be obtained using the Bogoliubov-de Gennes (BdG) equation~\cite{annett2004superconductivity}
\begin{equation}
E(q) = \varepsilon_q \sqrt{1+ 2 n v_{\rm RD}(q)/\varepsilon_q },
\end{equation}
where $v_{\rm RD}(q)$ is the Fourier transform of the bare interaction $v_{\rm RD}(r)$. At the mean-field (MF) level, the quasiparticle dispersion is given in terms of a single dimensionless parameter $\alpha^{\rm 3D}=nmUR_c^5/\hbar^2$ and $\alpha^{\rm 2D}=nmUR_c^4/\hbar^2$ in 3D and 2D, respectively~\cite{PhysRevLett.104.195302, macri2014ground}.
In the upper panel of Fig.~\ref{fig:E_MF} we compare the BdG excitation spectrum with the BF spectrum calculated from the HNC-EL/0 static structure factor, for different combinations of $\tilde{U}$ and $\tilde{R}_c$
such that $\alpha$ is fixed to $\alpha=30$. For large values of $\tilde{R}_c$, the more accurate
BF spectrum approaches the MF result. Overall the BdG mean field spectrum is quite adequate in 3D.  However, the
deviation of the MF results is substantially larger in 2D, because correlations are more important
in lower dimensions. In particular, the MF approximation does not predict the correct wave number of the roton, which the HNC-EL/0 results show to depend on $\tilde{R}_c$ and $\tilde{U}$ individually, and is not a universal function
of $\alpha$ only.

In the bottom panels of Fig. \ref{fig:E_MF}, we present the PDF $g(r)$ for fixed $\alpha$ and for different values of $\tilde R_c$ in three and two dimensional RDB gases, obtained within the HNC-EL/0 approximation. As for the comparison of the excitation spectrum, we vary
$\tilde{R}_c$ and $\tilde{U}$ for a fixed $\alpha=30$. $g(r)$ clearly depends not just on $\alpha$ but
on both $\tilde{R}_c$ and $\tilde{U}$. In both 3D and 2D, $g(r)$ is sensitive to the choice of
$\tilde{R}_c$ mostly for small $r$, which therefore cannot be described by the MF approximation.

\subsection{Off-diagonal long-range order and condensate fraction} \label{sec:1BDM-res}

We use the extension of the HNC-EL method to the one-body density matrix, summarized
in section~\ref{sec:1BDM}, to investigate the effects of interaction-induced correlations on
the off-diagonal long-range order and particularly on the condensate fraction.

\begin{figure}
\centering
\includegraphics[width=1.0\linewidth]{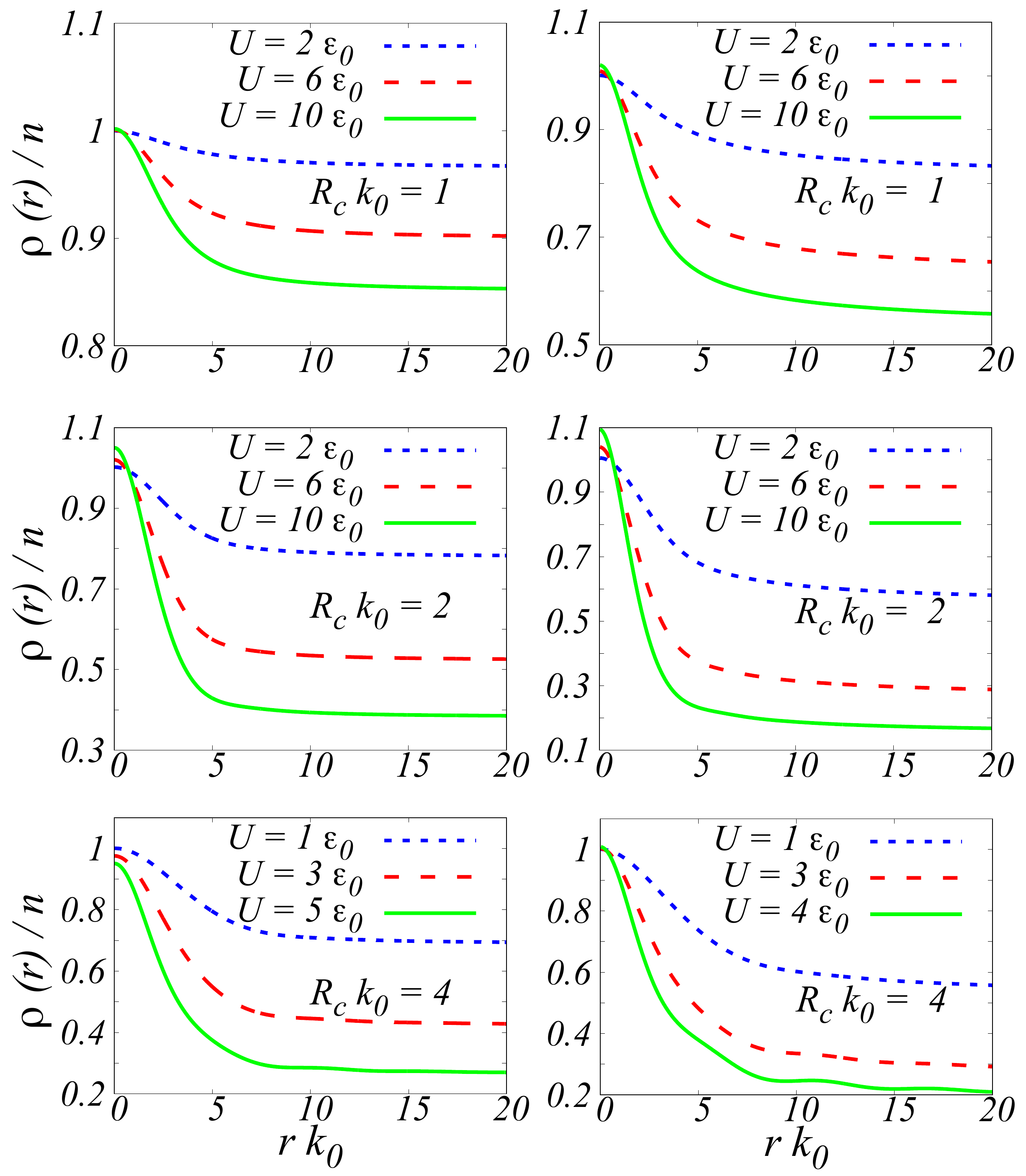}
\caption{The one-body density matrix $\rho (r)$ of a 3D (left) and 2D (right) gas of RDB versus $r k_0$ for different values of  $\tilde{R}_{c}$ and $\tilde{U}$.
\label{fig:rho}}
\end{figure}
\begin{figure}
\centering
\includegraphics[width=1.0\linewidth]{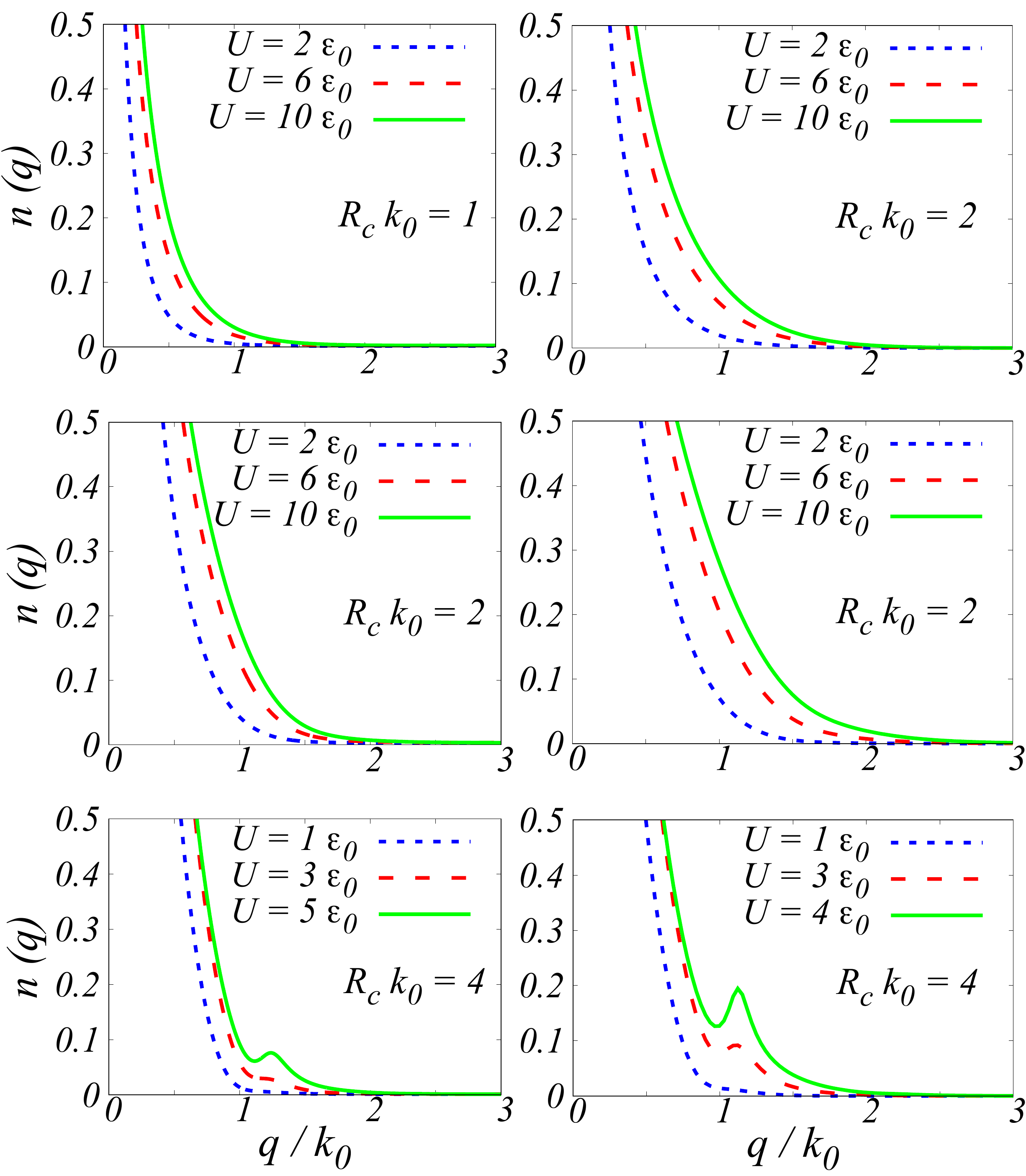}
\caption{The momentum distribution function $n (q)$ of a 3D (left) and a 2D (right) gas of RDB versus $q/k_0$ for different values of $\tilde{R}_{c}$ and $\tilde{U}$.
\label{fig:nk}}
\end{figure}

In Figs.~\ref{fig:rho} and \ref{fig:nk} we show our results for the OBDM and the momentum distribution function of RDB gases, respectively. We observe ODLRO, i.e. a non-zero limit of
$\rho(r)/n$ for $r\to\infty$, for all combinations of ${\tilde R}_c$ and ${\tilde U}$ that we studied, because
all HNC-EL/0 calculations are for the homogeneous gas of RDB. With increasing interaction strength ${\tilde U}$
and increasing soft-core radius ${\tilde R}_c$, the effect of ODLRO is suppressed as expected and seen by a
decreasing asymptote $\rho(r\to\infty)/n$.
The oscillatory behavior of the OBDM and a finite momentum peak in the momentum distribution function $n(q)$ of both 3D and 2D systems at large ${\tilde R}_c$ and ${\tilde U}$ are noticeable (see the bottom panels in Figs.~\ref{fig:rho} and \ref{fig:nk}).
Both of these features signal the tendency of a homogeneous superfluid towards the formation of inhomogeneous phases.
Hence a time-of-flight measurement of $n(q)$ could provide evidence for an instability against formation of a droplet
crystal phase, seen as a finite momentum peak in $n(q)$.
Also, notice that the unphysical divergence in the long wavelength limit of $n(q)$ has its roots in the failure of HNC-EL/0 approximation in reproducing the correct asymptotic behavior of OBDM at large distances~\cite{PhysRevB.31.7022} (see, Appendix~\ref{appendix1} for more details). The small deviation of $\rho(r)/n$ from the
exact value unity for $r=0$ is a gauge for the accuracy of the HNC-EL/0
approximation~\cite{krotscheckPRB85condensate}. While previous
studies of $^4$He were afflicted by a major deviation from unity, the deviation for the RDB is only a few
percent in the cases studied here, which indicates that HNC-EL/0 is sufficiently accurate for the RDB,
see also the comparison with the exact Monte Carlo results below.

\begin{figure}
\includegraphics[width=0.8\linewidth]{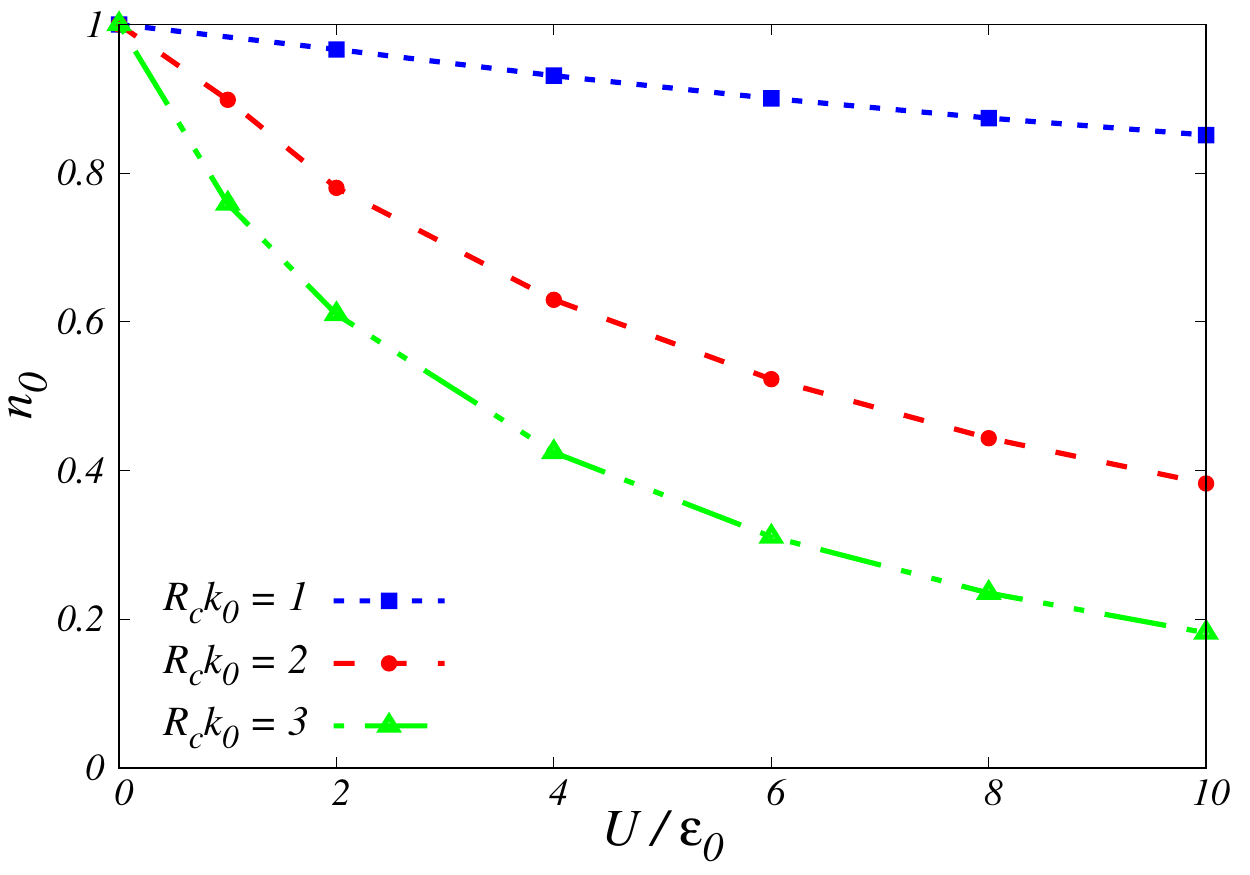}\\
\includegraphics[width=0.8\linewidth]{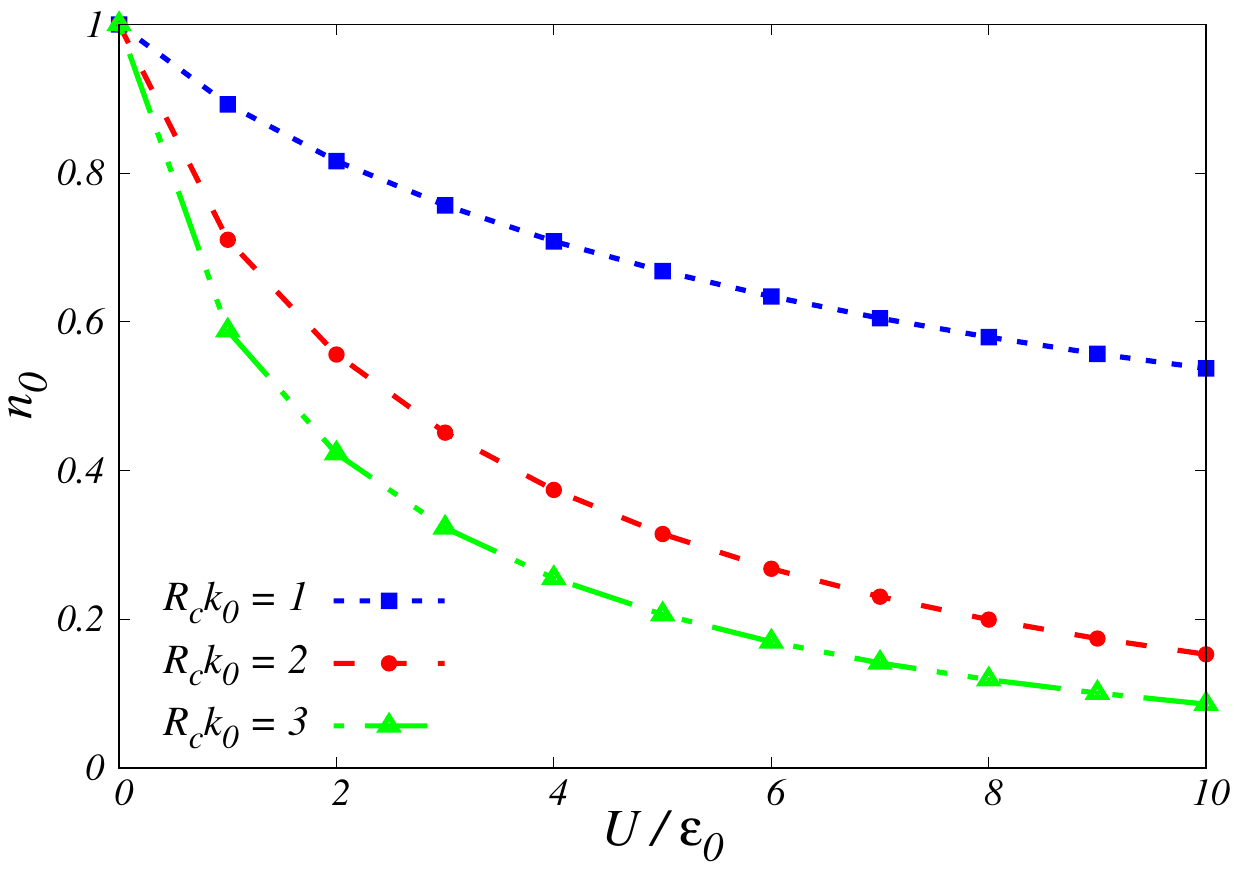}
\caption{The condensate fraction $n_0$ of a 3D (top) and 2D (bottom) gas of RDB as a function of $\tilde{U}$ for different values of $\tilde{R}_{c}$.
\label{fig:n0}}
\end{figure}

The asymptotic value of $\rho(r)/n$ for $r\to\infty$ is the condensate fraction $n_0$ and is
presented in Fig.~\ref{fig:n0}. 	As discussed above,
the condensate fraction decreases with increasing either the interaction strength ${\tilde U}$ or the soft-core radius ${\tilde R}_c$ but it remains finite even in the region where the homogeneous phase is only meta-stable. 

\subsection{The ground state energy}\label{sec:E_GS}
The ground state energy per particle within the HNC-EL/0 approximation is obtained from~\cite{Fabrocini2002book}
\begin{equation}
\begin{split}
\varepsilon^{\rm HNC}_{\rm GS} (\tilde{U} , \tilde{R}_c) =
& \frac{n}{2} \int \mathrm{d} \textbf{r}
\left[ g(r) v_{\rm RD}(r)+\frac{\hbar^2}{m} \left|\nabla \sqrt{g(r)}\right|^2 \right]\\
&- \frac{\hbar^2}{8m n} \int \frac{\mathrm{d} \textbf q}{(2 \pi)^D} \frac{q^2\left[S(q)-1\right]^3}{S(q)},
\end{split}
\end{equation}
in which many-body correlations beyond the mean-field level are approximately accounted for. 
The difference between the ground-state energy and the Hartree energy is conventionally called the correlation energy
$\varepsilon_c = \varepsilon_{\rm GS} - \varepsilon_{\rm H}$,
where the Hartree energy per-particle $\varepsilon_{\rm H}=n v_{\rm RD}(q=0)/2$ 
is given by $\varepsilon_{0} \tilde{U} \tilde{R}_c^3/18$ and
$\varepsilon_{0} \pi \tilde{U} \tilde{R}_{c}^2/(12 \sqrt{3})$, in 3 and 2 dimensions, respectively.
Note that, in the mean field approximation, the kinetic energy is zero for a homogeneous system.

\begin{figure}
\centering
\includegraphics[width=0.8\linewidth]{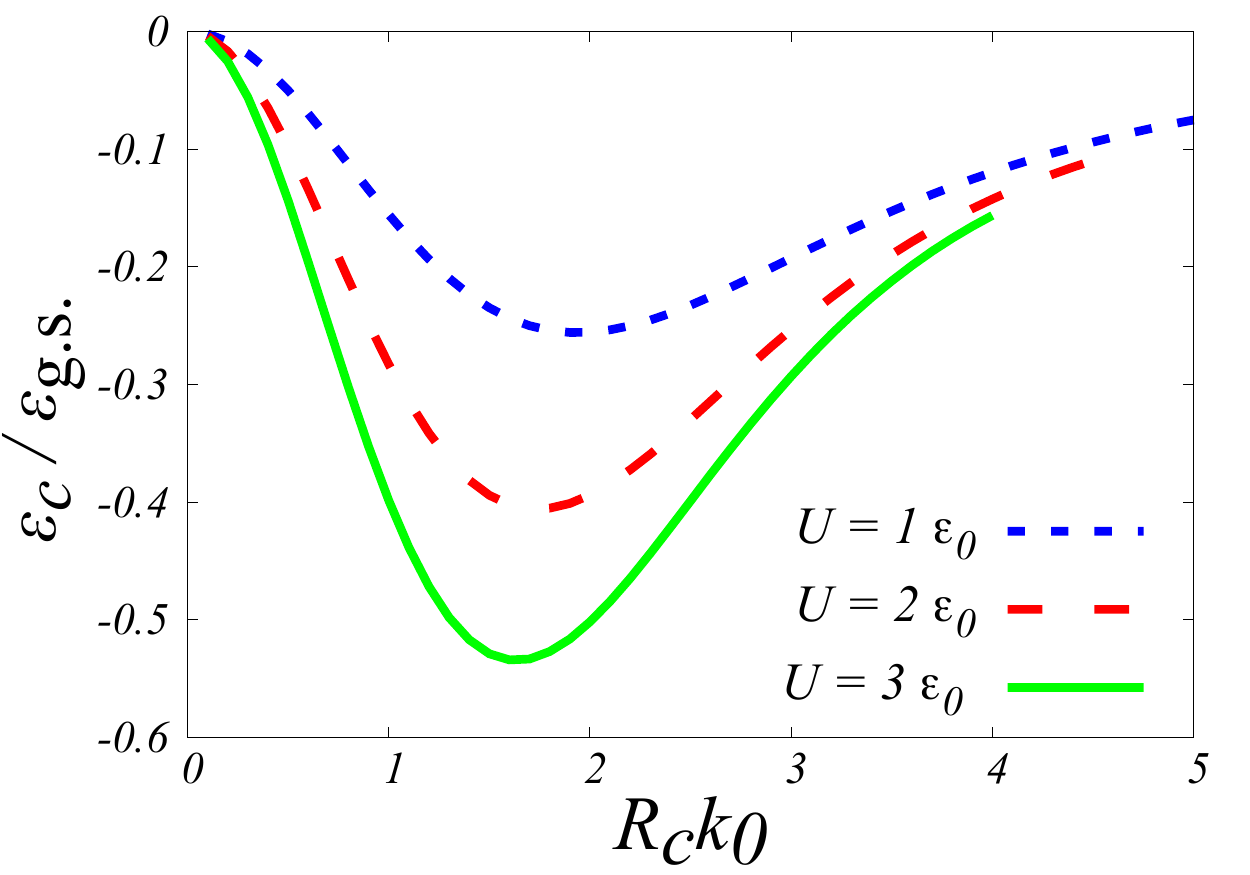}	\\
\includegraphics[width=0.8\linewidth]{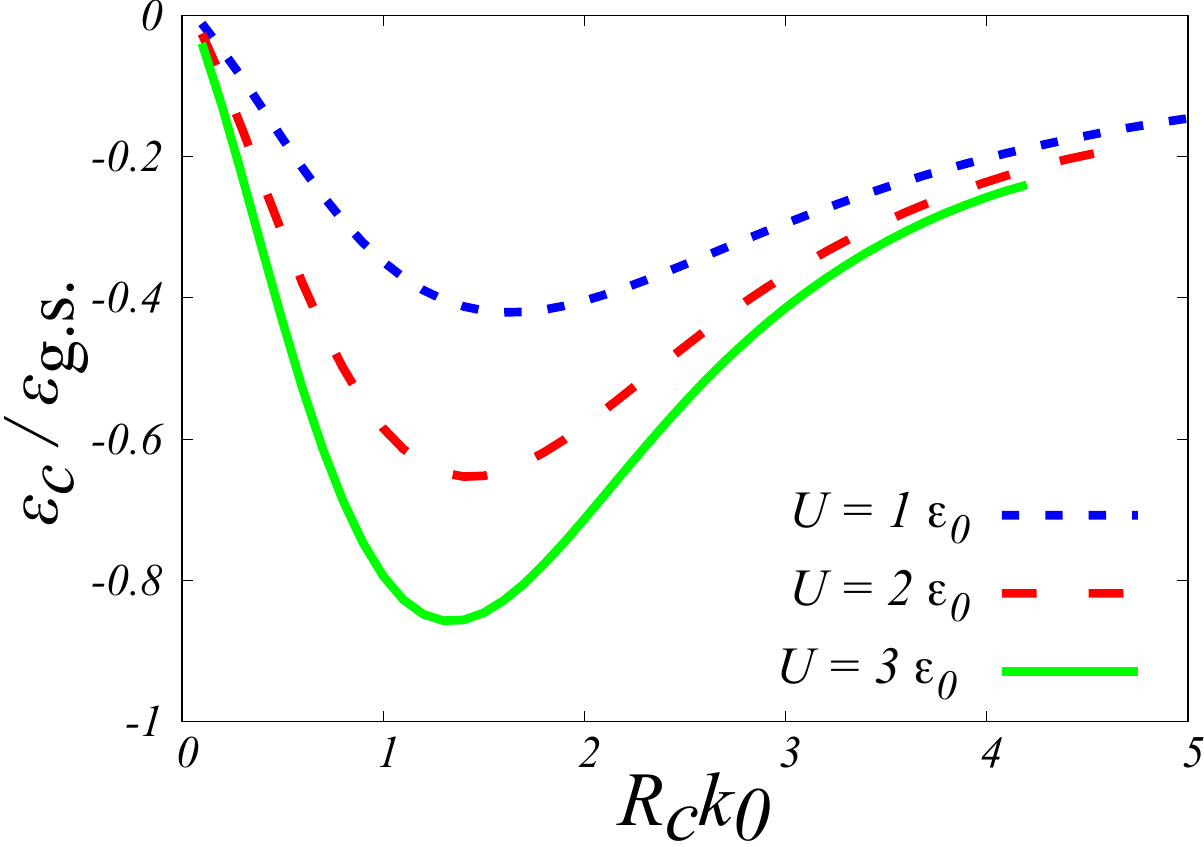}
\caption{The correlation energy per particle $\varepsilon_c$ of 3D (top) and 2D (bottom) RDB gas, in units of the ground state energy $\varepsilon_{\rm GS}$, as a function of the soft-core radius ${\tilde R}_c$ at several values of $\tilde{U}$ calculated within the HNC-EL/0 formalism.
\label{fig:E_Rc}}
\end{figure}

In Fig.~\ref{fig:E_Rc}, we report our numerical findings for the correlation energy $\varepsilon_c$
of a RDB gas within the HNC-EL/0 approximation.
As expected the correlation energy is negative, since the HNC-EL/0 method is based on a better variational ansatz -- the Jastrow-Feenberg ansatz -- than the mean field approximation, which lacks correlations. The correlation energy is comparable with the Hartree energy at intermediate values of the soft-core radii i.e., $R_c k_0 \simeq 1$.
While $\varepsilon_c$ increases monotonously with the interaction strength ${\tilde U}$, this is not the case for
its dependence on ${\tilde R}_c$: for both small and large values of ${\tilde R}_c$, the $\varepsilon_c$ becomes
negligible, and the HNC-EL/0 ground-state energy approaches the mean-field result.

\section{Monte Carlo simulations and transition to droplet crystal phase}\label{sec:PIMC}

For validation of the approximations used in the HNC-EL/0 calculations
(no elementary diagrams and no higher correlations than pair correlations),
we performed exact quantum Monte Carlo simulations. We used path integral Monte Carlo
(PIMC) simulations~\cite{chandlerJCP81} of $N=216$ Rydberg atoms in 3D with periodic boundary
conditions and including Bose symmetry. PIMC simulations yield unbiased and essentially exact results
for bosonic many-body systems and they are being widely used
for quantum fluids such as $^4$He~\cite{moroniJCP04,boninsegniPRE06,zillichJPC07,zillichJCP05}
and quantum gases~\cite{krauthPRL96,filinovPRA16}
including Rydberg gases \cite{PhysRevLett.105.135301,Cinti}.
The $N$-body density matrix is approximated by the pair
action, following Ref.~\cite{ceperley95}. This allows using fairly large
time steps $\tau$, reducing the path length and thus the computational
effort of our simulations.
Since PIMC simulates ensembles (the canonical ensemble in our
case) at finite temperature $T$, we reduced $T$ until the quantities
that we aim to compare, namely $g(r)$ and $S(k)$, become essentially independent
of $T$, which means the system is effectively in the ground state.
Details can be found in the appendix~\ref{appendix2}.
The properties of a thermal cloud of Rydberg atoms and the influence
of temperature on the transition to a crystalline phase would
constitute a separate investigation, but is not the subject of this work.

\begin{figure}
	\includegraphics[width=\linewidth]{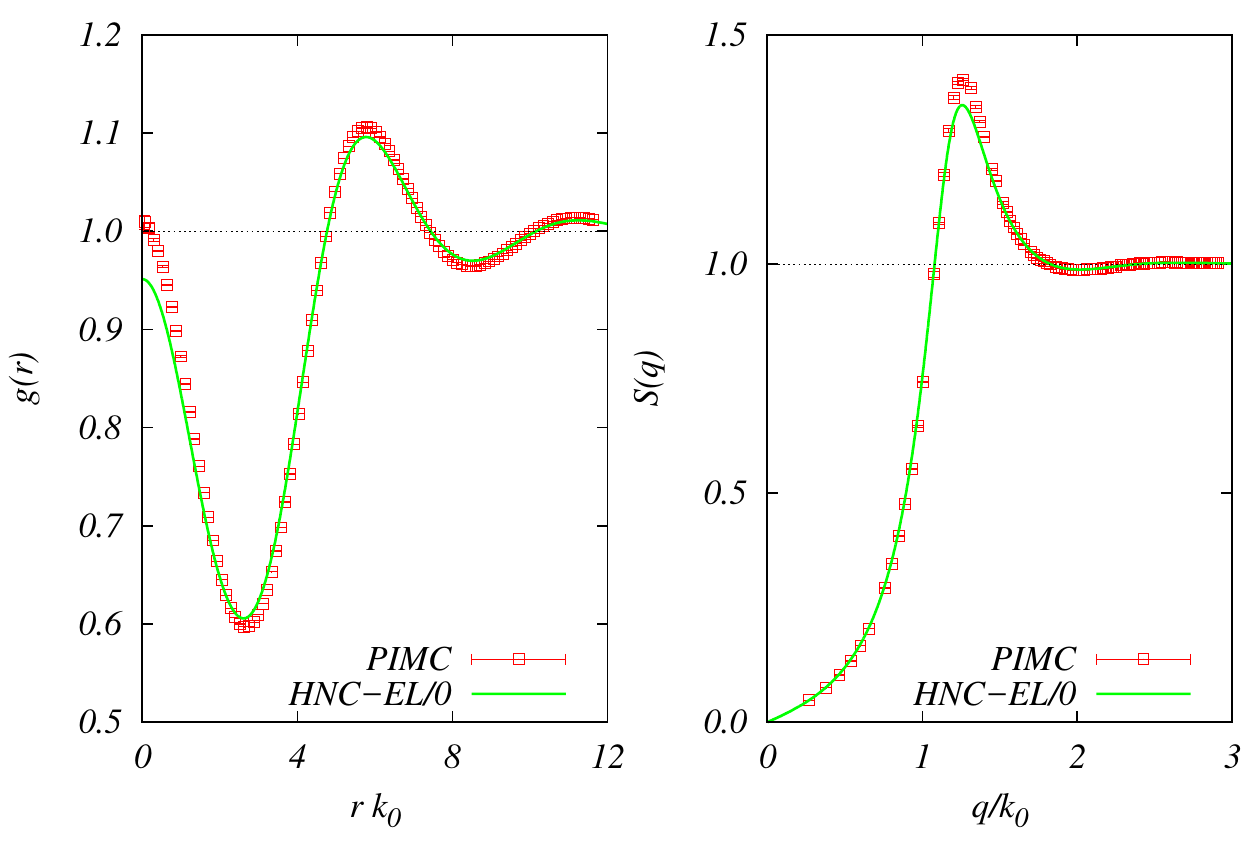}
	\caption[]{\label{FIG:grU3}
		Pair distribution function $g(r)$ (left) and static
		structure factor $S(k)$ (right) of a 3D gas of RDB for ${\tilde R}_c=4$ and ${\tilde U}=3.0$. The symbols show the
		PIMC results and the lines the HNC-EL/0 results.
	}
\end{figure}

In Fig.~\ref{FIG:grU3} we compare the HNC-EL/0 results for
$g(r)$ and $S(k)$ with the corresponding PIMC results, for ${\tilde R}_c=4$ and ${\tilde U}=3.0$.
The agreement is very good.  While HNC-EL/0 slightly underestimates the height of the main peak in $S(k)$, the peak position is extremely
well reproduced. This is important for estimating the lattice
constant of the self-assembled lattice in the density
wave state: the peak position of $S(k)$ for a fluid, i.e. homogeneous state
predicts the Bragg peak of the crystalline phase very well,
as we will see below. We note that for these values of ${\tilde R}_c$ and ${\tilde U}$ 
the PIMC results are independent of the starting
configurations of the Metropolis random walk simulating the
canonical ensemble.

\begin{figure}
	\includegraphics[width=0.9\linewidth]{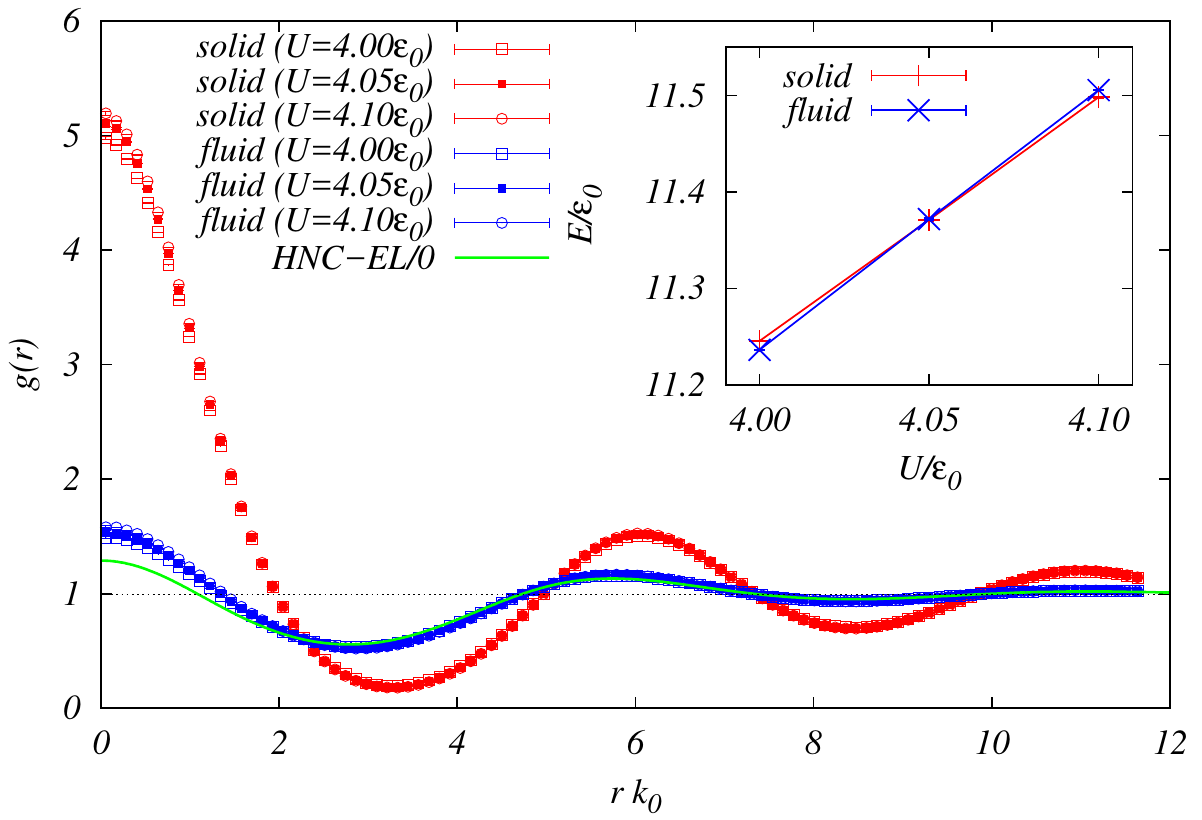}\\
		\includegraphics[width=0.92\linewidth]{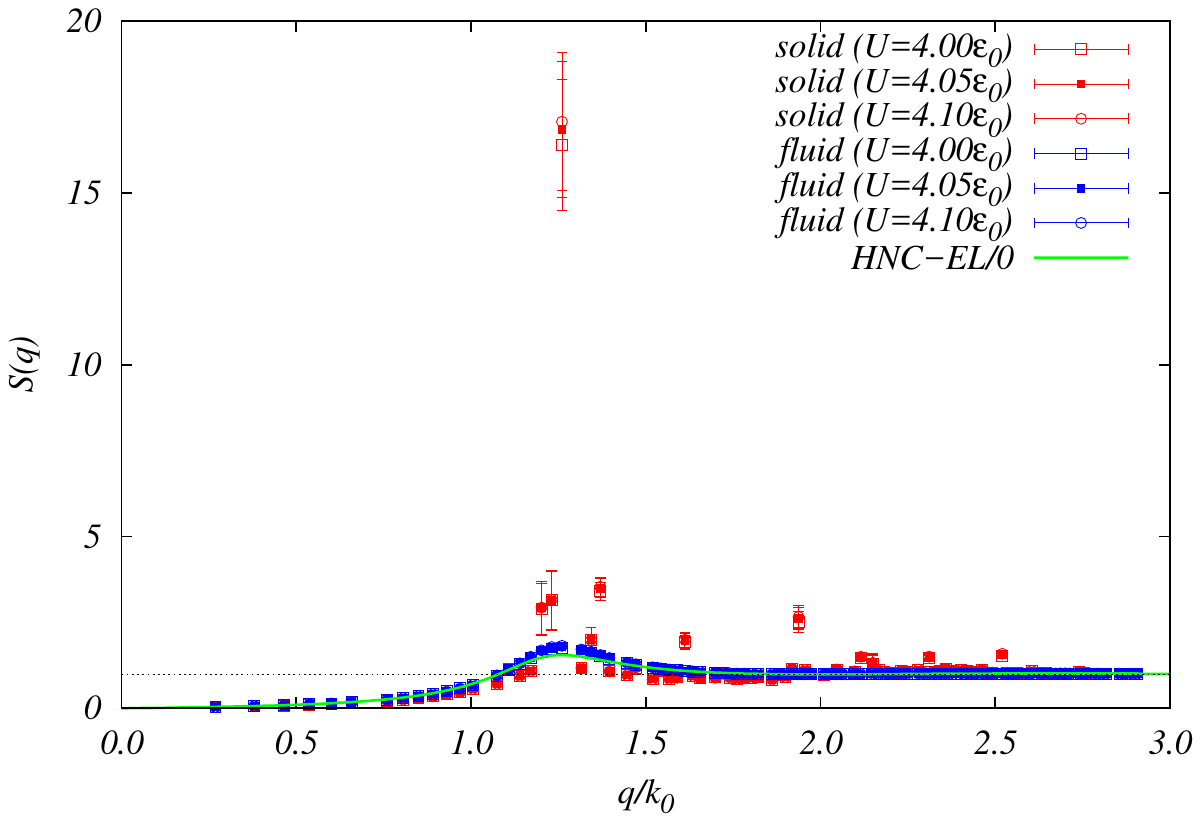}
	\caption[]{\label{FIG:grU4}
		(Color online) Top: Pair distribution function $g(r)$ of three-dimensional RDB for ${\tilde R}_c=4$ and ${\tilde U}=4.00;4.05;4.10$.
		The red symbols with error bars show the PIMC results for
		$g(r)$ for the three values of ${\tilde U}$ in the density-wave state
		corresponding to a fcc lattice; the blue symbols show $g(r)$ for the same ${\tilde U}$ values,
		but in the fluid, i.e. homogeneous state. Also shown
		is the HNC-EL/0 result for the homogeneous state (green line) for ${\tilde U}=4.00$.
		The inset shows the energy per particle as a function of ${\tilde U}$ for the two
		states, indicating a first-order transition around ${\tilde U}=4.05$.
		Bottom: Same as top panel for the static structure factor $S(k)$. }
\end{figure}

Since the HNC-EL/0 calculations above indicate that the Rydberg gas
becomes unstable against density oscillations as ${\tilde U}$ (or ${\tilde R}_c$)
is increased, we naturally performed PIMC simulations also for
larger values of ${\tilde U}$.  For example already for ${\tilde U}=5$ and ${\tilde R}_c=4$  we find that
starting at a homogenous phase (e.g. from simulations with
${\tilde U}=3$), the system eventually crystallizes into a more or less
regular face-centered cubic lattice.  Note that
for ${\tilde R}_c=4$ and ${\tilde U}=5$, our HNC-EL/0 calculations, which uses
a {\em homogeneous} Jastrow ansatz, still converges to a homogeneous state without problems.
Considering the good agreement for ${\tilde U}=3$
(Fig.~\ref{FIG:grU3}), it is unlikely that HNC-EL/0 would fail
for somewhat larger values of $\tilde U$.  Our HNC-EL/0 calculations are based
on a homogeneous, i.e. translationally invariant, wave function, hence we can
only get homogeneous solutions. These are only metastable if there is an
inhomogeneous droplet crystal solution of lower energy. The transition
is thus expected not to be continuous, but a first order transition.
This has indeed been found using the mean field approximation
in three dimensions~\cite{PhysRevLett.104.195302} and using PIMC simulations in two
dimensions~\cite{Cinti}.
First order transitions are usually studied with quantum Monte
Carlo methods that employ a trial wave functions by comparing energies
obtained with the different trial wave function, e.g. a homogeneous
Jastrow wave function (as we use for HNC-EL) and a trial wave function
appropriate for a solid, see e.g. Ref.~\cite{astraPRL07,moroni2014}.  In this work
we try to use PIMC, which is unbiased by a trial wave function, to
investigate the phase transition from a uniform fluid to a crystal state.
In the vicinity of ${\tilde U}=4$ (and still ${\tilde R}_c=4$), we found that our PIMC results depend on
the initial configuration; even long equilibration did not
lead to a phase change between uniform fluid and crystal.  Therefore
we study the Rydberg gas in the vicinity of ${\tilde U}=4$ by either initializing the simulations
with crystal configurations obtained for ${\tilde U}=5$ 
\footnote{only using those simulations that happen to crystallize in a perfect fcc lattice and discarding fcc
	lattices with defects},
 or by initializing with uniform configurations
from ${\tilde U}=3$.  The results for these two sets of simulations are shown
in Fig.~\ref{FIG:grU4} for a narrow range of
${\tilde U}$ values, ${\tilde U}=4.0;4.05;4.1$.
Both the pair distribution functions $g(r)$ and the static structure factors
$S(k)$ differ strongly between the fluid and the crystal case for a given ${\tilde U}$.
In the crystal phase $g(r)$ has a large peak at $r=0$, and falls
quickly to almost zero, followed by extended oscillations up to
the limit of half the box length. The corresponding peak in
$S(k)$ is evocative of the Bragg peak of a solid.

Conversely, the homogeneous fluid phase is characterized by a $g(r)$ with much weaker correlations
at $r=0$ and oscillations that decay much quicker to unity.  The corresponding
$S(k)$ has no Bragg peak but is a smooth function as expected for fluid states.
For ${\tilde U}=4.0$, the comparison between HNC-EL/0 (lines) and PIMC (blue symbols) still
shows good agreement, predicting the correct peak position in
$S(k)$. HNC-EL/0 exhibits weaker correlations in $g(r)$; this is the
usual consequence of the approximations made in HNC-EL/0, which
can be improved by including elementary diagrams and/or triplet
correlations at least approximately. We note that beyond
the first-order transition to a crystal -- at ${\tilde U} \approx 4.05$ in the
case of ${\tilde R}_c=4$ -- HNC-EL/0 still gives valid, albeit approximate results:
HNC-EL/0 based on a homogeneous Jastrow wave function explores the metastable
regime of the homogeneous fluid phase.  Expressed in terms of the dimensionless parameter
	$\alpha^{\rm 3D}$, our PIMC simulations predicts the phase transition to
	occur at $\alpha^{\rm 3D}=35$, which is slighly higher than the mean field
	estimate of $\alpha^{\rm 3D}=30$ \cite{PhysRevLett.104.195302}.

The energies of the crystal
and fluid phases are shown in the inset of Fig.~\ref{FIG:grU4}.
For example, the ground-state energy per-particle for ${\tilde U}=4.0$ and ${\tilde R}_c=4.0$ is
	$\varepsilon_{\rm GS}=11.25\, \varepsilon_0$, while the HNC-EL/0 result for this ${\tilde U}$ and ${\tilde R}_c$ is
	$\varepsilon_{\rm GS}=12.19 \, \varepsilon_0$ -- slightly higher as expected for a variational method.
	The PIMC energies of the crystal and fluid intersect at a critical ${\tilde U}_c\approx 4.05$.
 As expected, the energy of the fluid phase is lower for ${\tilde U}<{\tilde U}_c$ and the energy
of the crystal is lower for ${\tilde U}>{\tilde U}_c$. The crossing of the energies and
the behavior of $g(r)$ and $S(k)$ is a strong indicator for a first-order transition, which is not surprising for
	a liquid-solid transitions. Note however, that in the present
	case we have a quite peculiar solid \cite{PhysRevLett.104.195302,PhysRevLett.105.135301,Cinti}
	(which has been found also for classical systems with similar interactions\cite{liuPRE08,archerJPhysCondMat08}):
	a lattice site of this solid consists of a fluid cluster of atoms rather than of a single atom.  The droplet
	size $N_d$ depends on $U$ and $R_c$.  
For instance, for the parameters in Fig.~\ref{FIG:grU4}, the droplets consist of slightly less than $N_d=3$ particles on average.
This can be obtained by integrating the peak of $g(r)$ at $r=0$ up to the first minimum at $r_{\rm min}$,
$N_d=1+4\pi n \int_0^{r_{\rm min}} \mathrm{d}r\, r^2\, g(r)$.

\section{Summary}\label{sec:summ}

We have studied the ground-state properties of Rydberg-dressed Bose gases in two and three dimensions by means of hypernetted-chain approximation and, for quantitative comparison, path integral Monte Carlo simulations.
For a homogenous fluid, the HNC approximation even in its simplest level, i.e., HNC-EL/0 gives results in very good agreement with the PIMC data, while requiring orders of magnitude lower computational effort.
The pair distribution function and excitation spectrum signal the tendency of the
homogenous superfluid phase to become unstable against density waves when the interaction strength
$U$ or the soft-core radius $R_c$ are increased. Based on the HNC-EL/0 ground state structure function, we calculated the excitations spectra using the Bijl-Feynman approximation~\cite{{PhysRev.102.1189},{PhysRev.94.262}}.
Close to the instability, the excitation spectrum
exhibits a pronounced roton minimum, which is a precursor to establishing long-range order, i.e. crystallization.
The comparison of our results for the spectra with the mean field approximation showed that, for the excitation spectrum,
the MF approximation is adequate in 3D, but deviates significantly from our more accurate results in 2D. For other quantities, such as the pair distribution function for small pair distances or the ground state energy, the deviations of the MF approximation are significant also in 3D. In particular, we show that the spectrum does not
depend universally on a single parameter characterizing the Rydberg interaction but on the coupling strength and soft-core radius individually.  The PIMC simulations for 3D confirmed the
homogeneous phase undergoes a first order phase
transition to a droplet crystal phase \cite{PhysRevLett.104.195302,PhysRevLett.105.135301,Cinti}. 
At strong coupling strengths, when the soft-core radius of the interaction is comparable with the average inter-particle separation i.e., $R_c k_0\simeq 1$, the correlation energy becomes comparable with the Hartree, i.e.
mean-field energy, and strongly lowers the total ground state energy towards the exact value. We also studied off-diagonal long-range order
and found that the inter-particle interaction strongly depletes the Bose-Einstein condensation, but even in the vicinity of the transition to a droplet crystal, the condensate fraction remains finite.

For the calculation of the excitation spectrum, we used the simple Bijl-Feynman approximation, which
provides an upper bound to the true spectrum.  For example in $^4$He, the Bijl-Feynman approximation
overestimates the true roton energy by a factor of two. With improved methods, such as the correlated
basis method~\cite{Saarela86,krotscheckJCP01} or recent improvements thereof
\cite{campbellPRB09,campbellJLTP10,campbellPRB15}, nearly exacts can be obtained
for the excitation spectrum, including quantitative predictions for broadening due to damping. This will be
the topic of future work.

\acknowledgments{S.H.A. and R.A. are supported by Iran. BT is supported by TUBITAK and TUBA.
REZ acknowledges the support and facilities of
the Department for Scientific Computing at the Johannes Kepler University.
}

\appendix
\section{Long-wavelength behavior of the momentum distribution function}\label{appendix1}

\begin{figure}
\includegraphics[width=\linewidth]{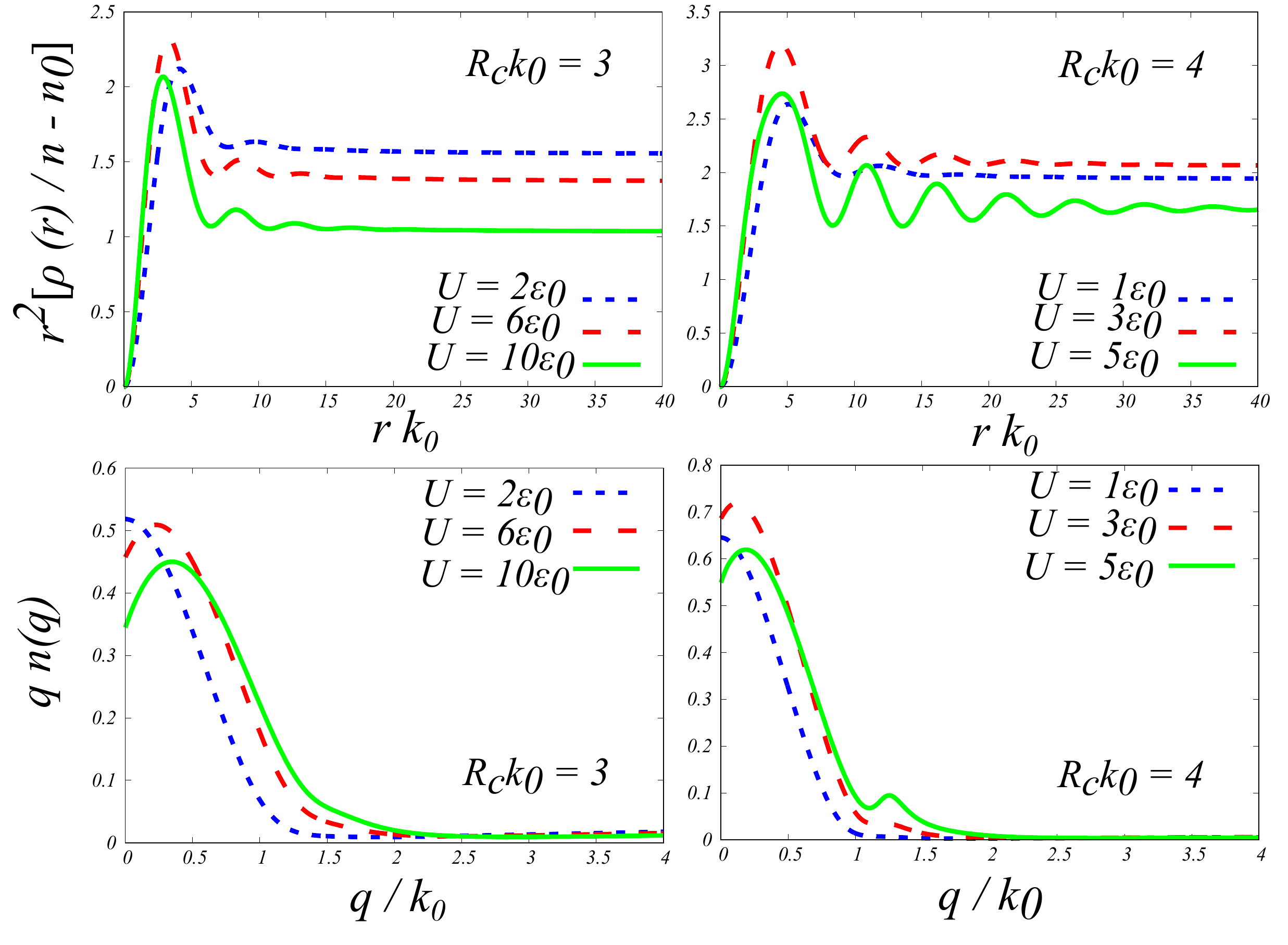}
\caption{Plots of $r^{2} [\rho(r)/n - n_0]$ (top) in units of $1/k_0^2$ and $q \, n (q)$ (bottom) in units of $k_0$ for a 3D gas of RDB at different values of  $\tilde{R}_{c}$  and $\tilde{U}$.
\label{fig:knk}}
\end{figure}

The OBDM obtained within the HNC-EL/0 formalism approaches its asymptotic value slower than what one would expect from the exact results. This results in an unphysical long wavelength divergence in $n(q)$~\cite{PhysRevB.31.7022}.
In particular for a 3D system we find
$\rho_{\rm HNC}(r\rightarrow \infty)-n n_0\propto 1/r^{2}$,
which results in
$n_{\rm HNC}(q\rightarrow 0)\propto 1/q$.
This clearly indicates an unphysical divergence in the HNC-EL/0 results for the momentum distribution function at long wavelengths.
Similar analyses of the numerical data in 2D gives $\rho( r \to\infty)-n n_0\propto 1/r^\gamma$ with $\gamma \approx 1.5$.
We have illustrated the behavior of $r^{2} [\rho(r) / n - n_0]$ and $q n(q)$ for a 3D RDB system in Fig.~\ref{fig:knk}, which better illustrates the oscillatory behavior of 1BDM, and the finite momentum peak of $n(q)$ at strong couplings.

We have illustrated the behavior of $r^{2} [\rho(r) / n - n_0]$ and $q n(q)$ for a 3D RDB system in Fig.~\ref{fig:knk}, which better illustrates the oscillatory behavior of 1BDM, and the finite momentum peak of $n(q)$ at strong couplings.

\section{PIMC simulation details}\label{appendix2}

We compared our zero-temperature
variational results using the HNC-EL method with
results obtained with PIMC of $N=216$ Rydberg atoms in a simulation
with periodic boundaries. In order to assess the accuracy of the
PIMC results, we have to ensure that {\em (i)} the time step bias
is negligibly small and that {\em (ii)} the temperature $T$ of the
PIMC simulations is chosen sufficiently small such that the
system is essentially in ground state.

\begin{figure}
\includegraphics[width=\linewidth]{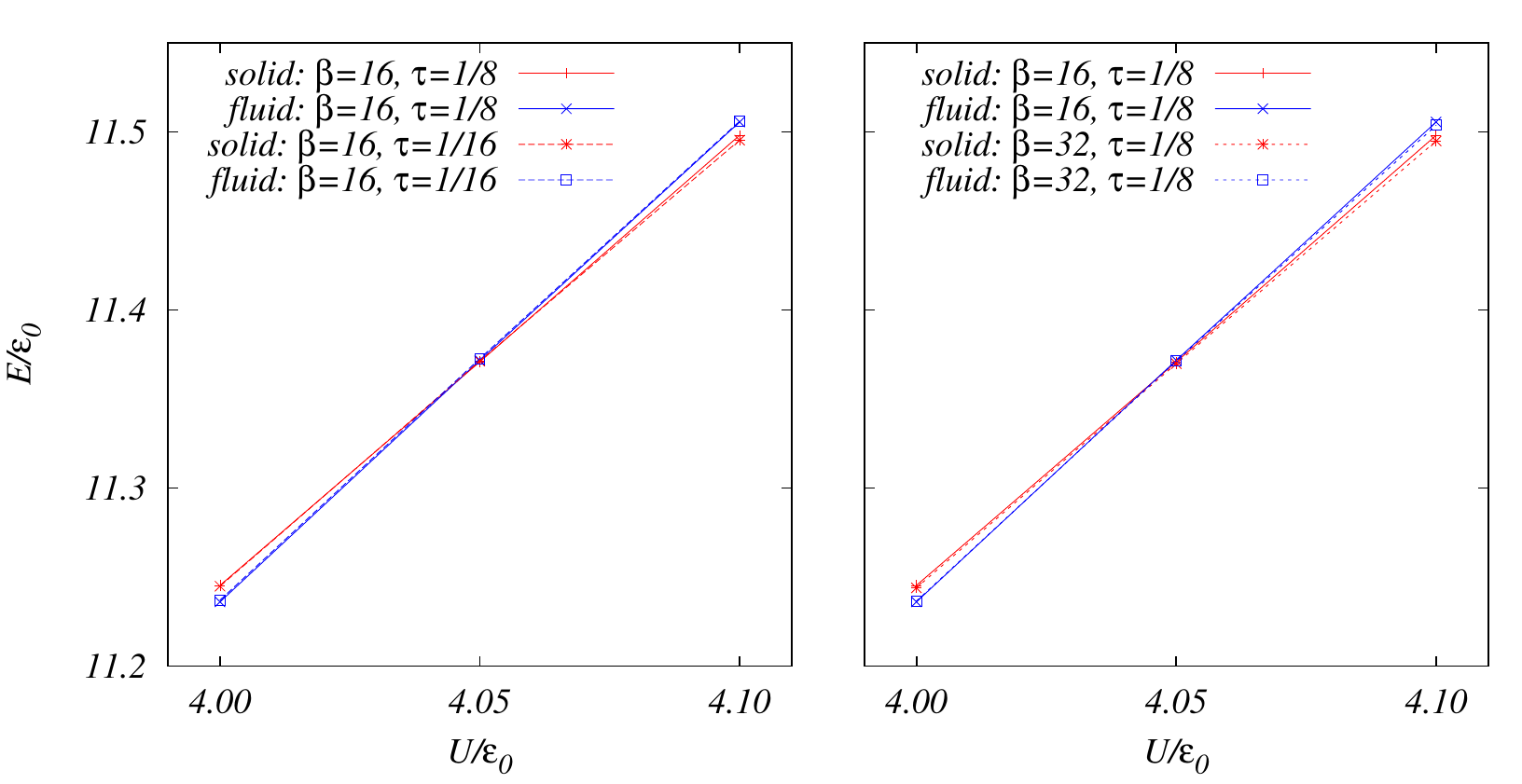}
\caption{The energies as shown in the inset of Fig.~\ref{FIG:grU3}
  for the fluid (blue) and solid (red) phase at inverse temperature
  $\beta=1/k_BT=16\,\varepsilon_0^{-1}$ and time step
  $\tau=1/8\, \varepsilon_0^{-1}$ are shown with full lines.
  Left: comparison with results at half the time step, i.e. $\tau=1/16\, \varepsilon_0^{-1}$,
  shown as dashed lines.
  right: comparison with results at half the temperature, i.e. $\beta=32\,\varepsilon_0^{-1}$,
  shown as dashed lines.  In all cases the error bars are smaller than the symbol size.
\label{fig:Ecomp2}}
\end{figure}

In our PIMC simulation we approximated the $N$-body density matrix
by a product of pair density matrices~\cite{ceperley95}, which at relatively
large imaginary time steps $\tau$ is much more accurate than the
Trotter approximation of the density matrix. For the results shown in
Figs.~\ref{FIG:grU3} and \ref{FIG:grU4}, we used time step $\tau=1/8\, \varepsilon_0^{-1}$,
at an inverse temperature $\beta=1/k_BT=16\,\varepsilon_0^{-1}$.
In Fig.~\ref{fig:Ecomp2} we demonstrate for $\tilde R_c=4$ that with these parameters our
PIMC simulations deliver essentially exact ground state ($T\to 0$) results.
The left panel compares the energies in the inset of Fig.~\ref{FIG:grU3}, shown
with full lines,
with energies obtained at half the time step, $\tau=1/16\, \varepsilon_0^{-1}$,
shown with dashed lines.
The right panel compares those energies with the result of a simulation
with half the temperature, i.e. twice the inverse temperature $\beta=32\,\varepsilon_0^{-1}$,
again showns as dashed lines. The full and dashed lines can hardly be distinguished,
demonstrating that our results are converged with respect to $\tau\to 0$ and
and $\beta\to\infty$. In all comparisons the phase transition occurs
at about $\tilde U=4.05$.

\begin{figure}
\includegraphics[width=\linewidth]{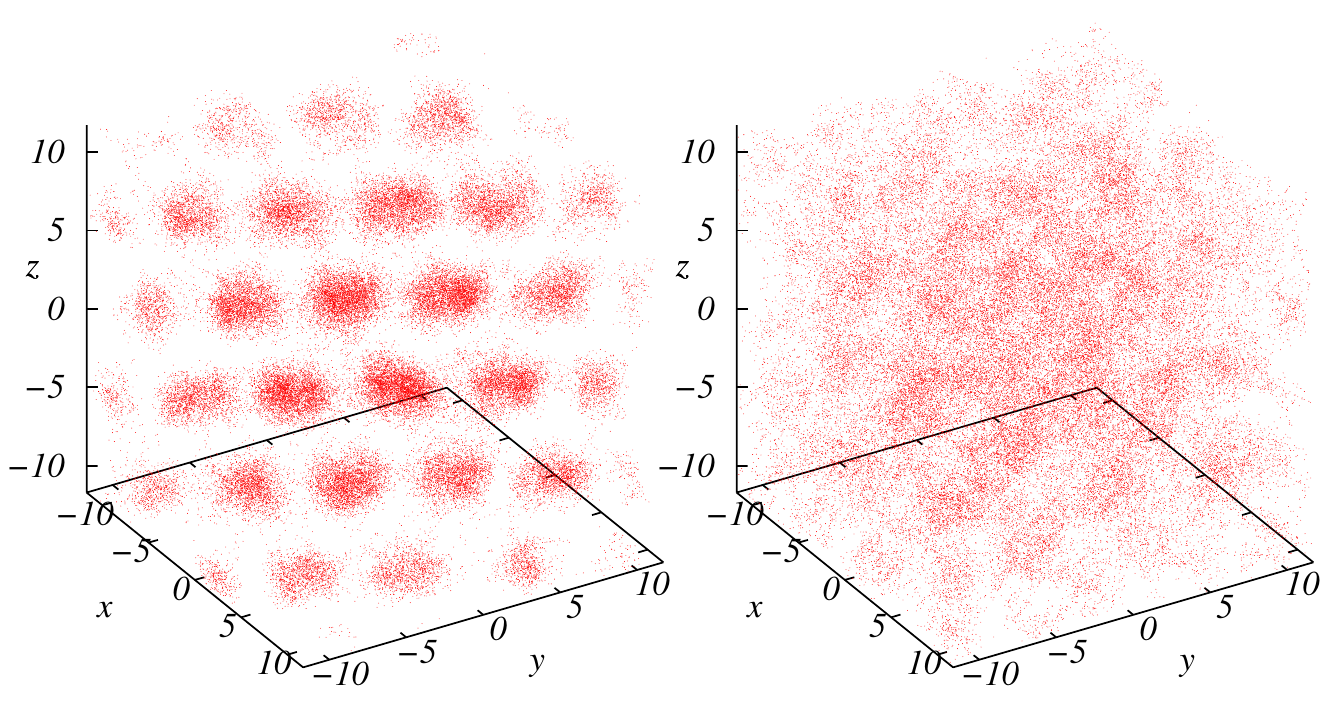}
\caption{PIMC simulation snapshots showing the beads of a simulation
  of the solid (left) and liquid (right) phase for  $\tilde U=4.1$
\label{fig:100x}}
\end{figure}

In order to illustrate the long range order of the solid phase, we show
a snapshot of the PIMC simulation of solid phase for $\tilde U=4.1$
and $R_c=4$
in the left panel of Fig.~\ref{fig:100x} where each bead of each Rydberg
atom is represented by a dot.  The triangular structure of the face-centered
cubic lattice when viewed along a diagonal of the cube is clearly visible.
The right panel is a snapshot of simulation of the (metastable) fluid
phase at the same interaction parameters, showing no long-range order.

\bibliography{RDBG_v14.bbl}

\end{document}